\documentclass{PoS}

\newcommand{\GeV}{{\ensuremath\rm \,GeV}}
\newcommand{\be}{\begin{equation}}
\newcommand{\ee}{\end{equation}}
\newcommand{\bea}{\begin{eqnarray}}
\newcommand{\eea}{\end{eqnarray}}

\newcommand{\doublet}[2]{ \left( \begin{array}{c}#1 \\ #2 \end{array}\right) }

\newcommand{\Et}{\cancel{E}_{T}}

\usepackage{tikz}
\usetikzlibrary{decorations.pathmorphing,decorations.markings}

\tikzset{
photon/.style={decorate, decoration={snake,amplitude=2pt, segment length=5pt}, draw=black},
particle/.style={draw=black, postaction={decorate}, decoration={markings,mark=at position .5 with {\arrow[draw=black]{>}}}},
antiparticle/.style={draw=black, postaction={decorate}, decoration={markings,mark=at position .5 with {\arrowreversed[draw=black]{>}}}},
gluon/.style={decorate, draw=black, decoration={coil,amplitude=4pt, segment length=5pt}},
goldstone/.style={draw=green,postaction={decorate},decoration={markings,mark=at position .5 with {\arrow[draw=blue]{>}}}}
}

\title{Dark Matter Signals at the LHC from a 3HDM}

\ShortTitle{DM signals at the LHC from a 3HDM}

\author{\speaker{Diana Rojas-Ciofalo}\\
        University of Southampton\\
        E-mail: \email{D.Rojas-Ciofalo@soton.ac.uk}}

\author{Adriana Cordero,$^a$ Jaime Hernandez-Sanchez,$^a$ Venus Keus,$^b$ Stefano Moretti,$^c$ Dorota Sokolowska$^d$ \\
        \llap{$^a$}Facultad de Ciencias de la Electronica, Benemerita Universidad Autonoma de Puebla,\\
        Apdo. Postal 542, C.P. 72570 Puebla, Puebla, Mexico\\
        \llap{$^b$}Department of Physics and Helsinki Institute of Physics,\\
        Gustaf Hallstromin katu 2, FIN-00014 University of Helsinki, Finland\\
        \llap{$^c$}School of Physics and Astronomy, University of Southampton,\\
        Southampton, SO17 1BJ, United Kingdom\\
        \llap{$^d$}International Institute of Physics, Universidade Federal do Rio Grande do Norte,\\
        Campus Universitario, Lagoa Nova, Natal-RN 59078-970, Brazil\\
        E-mail: \email{adriana.cordero@correo.buap.mx}, \email{jaime.hernandez@correo.buap.mx}, \email{Venus.Keus@helsinki.fi}, \email{S.Moretti@soton.ac.uk}, \email{dsokolowska@iip.ufrn.br}}
        
\abstract{We analyse new signals of Dark Matter (DM) at the Large Hadron Collider (LHC) in a 3-Higgs Doublet Model (3HDM) where only one doublet acquires a Vacuum Expectation Value (VEV), preserving a parity $Z_2$. The other two doublets are inert and do not develop a VEV, leading to a dark scalar sector controlled by $Z_2$, with the lightest CP-even dark scalar $H_1$ being the DM candidate. This leads to the loop induced decay of the next-to-lightest scalar, $H_2\to H_1\bar{f}f$ ($f=u,d,c,s,b,e,\mu,\tau$), mediated by both dark CP-odd and charged scalars. This is a smoking-gun signal of the 3HDM since it is not allowed in the 2HDM with one inert doublet and is expected to be important when $H_2$ and $H_1$ are close in mass. In practice, this signature can be observed in the cascade decay of the SM-like Higgs boson, $h\to H_1H_2\to H_1H_1\bar{f}f$ into two DM particles and di-leptons/di-jets, where $h$ is produced from either gluon-gluon Fusion (ggF) or Vector Boson Fusion (VBF). However, this signal competes with the tree-level channel $q\bar{q}\to H_1H_1Z^∗\to H_1H_1\bar{f}f$. We devise some benchmarks, compliant with collider, DM and cosmological data, for which the interplay between these modes is discussed. In particular, we show that the resulting detector signature, with missing energy and invariant mass of $\bar{f}f$ much smaller than $m_Z$, can potentially be extracted already during Run 2 and 3. For example, the $H_2\to H_1\gamma^{∗}$ and $\gamma^{∗}\to e^+e^−$ case will give a spectacular QED mono-shower signal.}

\FullConference{7th Annual Conference on Large Hadron Collider Physics - LHCP2019\\
		20-25 May, 2019\\
		Puebla, Mexico}

\begin{document}

\section{Introduction}

The Higgs mechanism of Electro-Weak Symmetry Breaking (EWSB) seems to be the one chosen by Nature to assign mass to fermions and weak gauge bosons. In its minimal realization, through a single Higgs {\it doublet}, it implies the existence of a single Higgs boson,
as discovered in 2012 at the Large Hadron Collider (LHC). Indeed, such a minimal 
Standard Model (SM) is compatible with a myriad of experimental results.
However, the well-known unanswered questions such as the origin of flavour, with three families of quarks and leptons,
as well as Dark Matter (DM), suggest that some extension beyond the SM (BSM) is necessary.

Given the existence of three families of quarks and leptons, it is not so far-fetched to imagine that there might also be three
families of Higgs doublets, where, as for the fermions, the replication is not prescribed by the SM gauge group. 
Indeed, it is possible that the three families of quarks and leptons could be described by the same symmetries that describe the three Higgs doublets. In such scenarios, this generation/family symmetry could be spontaneously broken along with the EW  symmetry, although some remnant subgroup could survive, thereby stabilizing a possible scalar DM candidate.
For certain symmetries, it is finally possible to find a Vacuum Expectation Value (VEV) alignment that respects the original symmetry of the potential which will then be responsible for the stabilization of the DM candidate.
In such 3-Higgs-Doublet Models (3HDMs), amongst the various symmetries which can govern them \cite{Ivanov:2011ae}--\cite{Keus:2013hya}, 
a simple possibility is a single $Z_2$, referred to here as Higgs parity,
which can prevent Flavour Changing Neutral Currents (FCNCs) and possible charge breaking vacua.  

In the present paper, we shall focus on the phenomenology of one of these 3HDMs, namely, the one
in which the third scalar doublet is even and the first and second inert\footnote{A doublet is termed ``inert'', or at times ``dark" or simply ``scalar'',  since it does not develop a VEV, nor does it couple to fermions.} doublets are odd
under the $Z_2$ parity.
We assume a vacuum alignment in the 3HDM space of $(0,0,v)$ that preserves the $Z_2$ symmetry (i.e., the Higgs parity).
Thus, we are led to consider a model with two inert doublets plus one Higgs doublet (I(2+1)HDM). This model may be regarded as an extension of the model with one inert doublet plus one Higgs doublet (I(1+1)HDM)\footnote{This model is known in the literature as the Inert Doublet Model (IDM), herein, we refer to it as I(1+1)HDM, thus clarifying the number of inert and active Higgs doublets.} proposed in 1976 \cite{Deshpande:1977rw} and studied extensively for the last few years (see, e.g., \cite{Ma:2006km}--\cite{LopezHonorez:2006gr}), by the addition of an extra inert scalar doublet.  The lightest neutral scalar or pseudoscalar field amongst the two inert doublets,
which are odd under the $Z_2$ parity, provides a 
viable DM candidate which is stabilized by the conserved $Z_2$ symmetry, displaying different phenomenological characteristics from the candidate emerging from the I(1+1)HDM case \cite{Grzadkowski:2010au}. Within this framework, we study some new SM-like Higgs decay channels offered by the extra inert fields, with the intent of isolating those which would enable one to distinguish between the I(2+1)HDM and I(1+1)HDM, assuming CP conservation throughout. 

In particular, we shall focus on the loop induced decay of the 
next-to-lightest scalar, $H_2 \to H_1 f \bar f$ ($f=u,d,c,s,b,e,\mu,\tau$),
mediated by loops involving both dark CP-odd and charged scalars.
This decay chain occurs in the I(2+1)HDM but not in the I(1+1)HDM, so it enables the two models to be distinguished.
In practice, the loop decay can be observed
in the cascade decay of the 
SM-like Higgs boson into two DM particles and a fermion-antifermion pair, 
$h\to H_1 H_2\to H_1 H_1 f \bar f$, wherein the $h$ state is produced from gluon-gluon Fusion (ggF) (i.e., $gg\to h$) or
Vector Boson Fusion (VBF) (i.e., $ q q^{(')}\to q q^{(')} h$). Notice,  however, that this mode competes with the 
tree-level channel $q\bar q\to H_1H_1Z^*\to H_1H_1f \bar f$
present also in the I(1+1)HDM. 
The resulting detector signature, $\Et\; f \bar f$, with the $ f \bar f$ invariant mass well below the
$Z$ mass, would indicate the presence of such a loop decay onset by a
small difference between $H_2$ and $H_1$ which would in turn identify a region of I(2+1)HDM parameter space largely precluded to the  tree-level process. Indeed, we will show that such a distinctive signature can possibly be extracted  at the LHC during Run 2 and/or Run 3. In fact, amongst the possible $f\bar f$ cases, a particularly spectacular one would be the one in which an electron-positron  pair is produced, eventually yielding an isolated mono-shower signal of
QED nature, owing to the fact that the dominant component (over the box topologies) of the loop signal is the $H_2\to H_1\gamma^*$ one, where the photon is (necessarily, because of spin conservation) off-shell, yet eventually  producing the $e^+e^-$ pair  in configurations where the fermions are soft and/or collinear. In assessing the
scope of the LHC in accessing this phenomenology, we  shall consider all  available theoretical
\cite{Keus:2014isa,Moretti:2015cwa} and 
experimental constraints \cite{Keus:2014jha,Keus:2015xya,Ade:2015xua,Aprile:2017iyp,Ahnen:2016qkx,Khachatryan:2016ctc,Khachatryan:2016vau} affecting the I(2+1)HDM parameter space, so as to eventually define some benchmark scenarios which can be tested at the  CERN machine.

The layout of the paper is as follows. In the  next section we describe the CPC I(2+1)HDM. In Sect.~\ref{cascades}, we introduce and discuss the aforementioned loop cascade decays. In Sect.~\ref{sec-loop} we perform 
all necessary calculations, both at tree and loop level, including analytic formulae for the $H_2 \to H_1 f\bar f$ case. 
In Sect.~\ref{results}, we present our results. We then conclude in Sect.~\ref{summa}. This work is fully based on \cite{Cordero:2017owj}.

\section{The model}
\label{3HDM}



The potential symmetric under the $Z_2$ symmetry can be written as:
\begin{eqnarray}
 V  &=& V_0 + V_{Z_2}, \\
V_0 &=& - \mu^2_{1} (\phi_1^\dagger \phi_1) -\mu^2_2 (\phi_2^\dagger \phi_2) - \mu^2_3(\phi_3^\dagger \phi_3) + \lambda_{11} (\phi_1^\dagger \phi_1)^2+ \lambda_{22} (\phi_2^\dagger \phi_2)^2  + \lambda_{33} (\phi_3^\dagger \phi_3)^2 \\
&& + \lambda_{12}  (\phi_1^\dagger \phi_1)(\phi_2^\dagger \phi_2)
 + \lambda_{23}  (\phi_2^\dagger \phi_2)(\phi_3^\dagger \phi_3) + \lambda_{31} (\phi_3^\dagger \phi_3)(\phi_1^\dagger \phi_1) \nonumber\\
&& + \lambda'_{12} (\phi_1^\dagger \phi_2)(\phi_2^\dagger \phi_1) 
 + \lambda'_{23} (\phi_2^\dagger \phi_3)(\phi_3^\dagger \phi_2) + \lambda'_{31} (\phi_3^\dagger \phi_1)(\phi_1^\dagger \phi_3),  \nonumber\\
V_{Z_2} &=& -\mu^2_{12}(\phi_1^\dagger\phi_2)+  \lambda_{1}(\phi_1^\dagger\phi_2)^2 + \lambda_2(\phi_2^\dagger\phi_3)^2 + \lambda_3(\phi_3^\dagger\phi_1)^2  + {\rm h.c.} 
\end{eqnarray}
This potential has only a $Z_2$ symmetry and no larger accidental symmetry\footnote{Note that adding extra $Z_2$-respecting terms, $(\phi_3^\dagger\phi_1)(\phi_2^\dagger\phi_3)$,  $(\phi_1^\dagger\phi_2)(\phi_3^\dagger\phi_3)$,  $(\phi_1^\dagger\phi_2)(\phi_1^\dagger\phi_1)$,  $(\phi_1^\dagger\phi_2)(\phi_2^\dagger\phi_2)$,  
does not change the phenomenology of the model. The coefficients of these terms, therefore, have been set to zero for simplicity. }.
We shall not consider CP-Violation (CPV) here, therefore we require all parameters of the potential to be real.

The minimum of the potential is realised for the following point:
\begin{equation}
 \phi_1= \doublet{$\begin{scriptsize}$ \phi^+_1 $\end{scriptsize}$}{\frac{H^0_1+iA^0_1}{\sqrt{2}}},\quad 
\phi_2= \doublet{$\begin{scriptsize}$ \phi^+_2 $\end{scriptsize}$}{\frac{H^0_2+iA^0_2}{\sqrt{2}}}, \quad 
\phi_3= \doublet{$\begin{scriptsize}$ G^+ $\end{scriptsize}$}{\frac{v+h+iG^0}{\sqrt{2}}}, 
\label{explicit-fields}
\end{equation}
with $v^2=\mu^2_3/\lambda_{33}$.

The doublets $\phi_{1}$ and $\phi_{2}$ are the inert doublets (with an odd charge under $Z_{2}$), whereas $\phi_{3}$ is the active (SM) doublet (with even charge under $Z_2$, same as all the SM particles). The field $h$ corresponds then to the SM Higgs with $m^{2}_{h} = 2\mu^{2}_{3} = (125 \GeV)^{2}$.

We can separate the inert particles into two families, or generations, with the second generation being heavier than the first generation.  We will refer to $(H_1,A_1,$ $H^\pm_1)$ as the fields from the first generation and to
$(H_2,A_2,H^\pm_2)$ as the fields from the second generation.

Each of the four neutral particles could, in principle, be the DM candidate, provided it is lighter than the other neutral  states. In what follows, without loss of generality, we assume the CP-even neutral particle $H_1$ 
from the first generation to be lighter than all other inert particles, that is:
\begin{equation}
m_{H_1} < m_{H_2}, m_{A_{1,2}},m_{H^\pm_{1,2}}.
\end{equation}
In the remainder of the paper the notations $H_1$ and DM particle will be used interchangeably and so will be their properties, e.g.,  $m_{H_1}$ and $m_{\rm DM}$.


We focus on a simplified case in where:
\be 
\label{lambda-assumption} 
\mu^2_1 = n \mu^2_2, \quad \lambda_3 = n \lambda_2, \quad \lambda_{31} = n \lambda_{23}, \quad \lambda_{31}' = n \lambda_{23}', 
\ee
The motivation for this simplified scenario is that in the $n=0$ limit the model reduces to the well-known I(1+1)HDM. We assume no specific relation among the other parameters of the potential.

With this simplification, it is possible to obtain analytical formulae for the parameters of the potential in terms of chosen physical parameters. In this study, we choose the set $(m_{H_1}, m_{H_2}, g_{H_1 H_1 h}$, $\theta_a, \theta_c, n)$ as the input parameters where $g_{H_1 H_1 h}$ is the Higgs-DM coupling. The meaningful parameters of the model are then defined as follows:
\bea
&& \mu_2^2 = \Lambda_{\phi_2} - \frac{m_{H_1}^2+m_{H_2}^2}{1+n},
\\
&& \mu_{12}^2 = \frac{1}{2} \sqrt{(m_{H_1}^2-m_{H_2}^2)^2 - (-1+n)^2 (\Lambda_{\phi_2} - \mu_2^2)^2 },
\\
&& \lambda_2 = \frac{1}{2v^2} (\Lambda_{\phi_2}  - \Lambda''_{\phi_2} ),\\
&& \lambda_{23} = \frac{2}{v^2} \Lambda'_{\phi_2}, 
\\
&& \lambda'_{23} = \frac{1}{v^2} (\Lambda_{\phi_2}  + \Lambda''_{\phi_2} - 2 \Lambda'_{\phi_2} ),
\\
&& \Lambda_{\phi_2} = \frac{v^2 g_{H_1 H_1 h}}{4(\sin^2 \theta_h + n \cos^2 \theta_h)},\\
&& \Lambda'_{\phi_2} = \frac{2 \mu_{12}^2}{(1-n) \tan 2 \theta_c},
\\
&& \Lambda''_{\phi_2} = \frac{2 \mu_{12}^2}{(1-n) \tan 2 \theta_a }.
\eea
The mixing angle in the CP-even sector, $\theta_h$, is given by the masses of $H_1$ and $H_2$ and the dark hierarchy parameter $n$:
\be
\tan^2 \theta_h = \frac{m_{H_1}^2 - n m_{H_2}^2}{n m_{H_1}^2 - m_{H_2}^2}.
\ee
Notice that we restore the $n=1$ limit of dark democracy discussed in \cite{Keus:2014jha, Keus:2015xya,Cordero-Cid:2016krd} with $\theta_h = \pi/4$. For the correct definition of $\tan^2 \theta_h$, the following two relations need to be satisfied: $m_{H_1}^2 < n m_{H_2}^2$ and $m_{H_1}^2 < \frac{1}{n} m_{H_2}^2$. 
Without loss of generality, we can limit ourselves to $n<1$, which will correspond to $\tan2\theta>0$ for $\theta_h < \pi/4$. Reaching other values of $n$ is a matter of reparametrisation of the potential.

\section{Decays at the LHC}
\label{cascades}

In the model studied here, there is one absolutely stable particle, $H_1$, as its decays into  SM particles are forbidden by the conservation of the $Z_2$ symmetry. By construction, all other inert particles, which are also odd under the $Z_2$ symmetry, are heavier than $H_1$ and hence unstable. The decays of these heavier inert particles may provide striking experimental signals for the I(2+1)HDM.

Access to the inert sector can be obtained through the SM-like Higgs particle, $h$,  and/or the massive gauge bosons, $Z$ and $W^\pm$, with the heavy inert particle subsequently decaying into $H_1$ and on- or off-shell $W^\pm/Z/\gamma$ states. In fact, in this model, $h$ can decay into various pairs of inert particles, leading to different signatures. We will consider here $h\to H_2 H_1$ decays. In such a case, as intimated, we will consider Higgs production at the LHC  through ggF and VBF.  

The interesting  production and decay patterns may occur both at tree- and loop-level. In the former case,  the colliding protons produce an off-shell gauge boson $Z^*$, which can, in turn, give us a $H_1 A_i$ pair ($i=1,2$), followed by the decay of $A_i$ into $H_1 Z^{(*)}\to H_1 f \bar f$. In the latter case, one would produce a $h$ state decaying into $H_1 H_2\to H_1H_1f \bar f$, via the loop decay $H_2 \to H_1 f \bar f$. 
For both cases, one ends up with a $\Et  f \bar f$ signature (possibly accompanied by a resolved forward and/or backward jet in case of VBF and an unresolved one in  ggF), i.e., a di-lepton/di-jet pair, which would generally be captured by the detectors,
alongside  missing transverse energy, $\Et$, induced by the DM pair.
Here, $f=u,d,c,s,b,e,\mu,\tau$. For the
cases in which the mass difference $m_{A_i}-m_{H_1}$ or $m_{H_2}-m_{H_1}$ is small enough (i.e., $\approx 2m_e$), only the electron-positron signature would emerge, thus leading to the discussed Electro-Magnetic (EM) shower. 

It is important here to notice that the loop decay  chain initiated by $h\to H_1 H_2$ is specific to the I(2+1)HDM case, while the one induced by $A_1\to H_1 Z^{(*)}$ may also pertain to the I(1+1)HDM case. (In fact, neither $H_2$ nor $A_2$ exists in the I(1+1)HDM, unlike $A_1$.) Moreover, when the decays are non-resonant, there is no way of separating the two $A_i$ ($i=1,2$) patterns.
In contrast, the extraction and observation of the decay $h\to H_1 H_2$ (followed by the loop decay $H_2  \to H_1 f \bar f$)
would represent clear evidence of the I(2+1)HDM. 
 
In the upcoming subsections, we will discuss the aforementioned tree- and loop-level decay modes of inert states into the DM candidate in all generality, then we will  dwell on the features of the $\Et  f \bar f$ signature.

\subsection{Tree-level decays of heavy inert states}

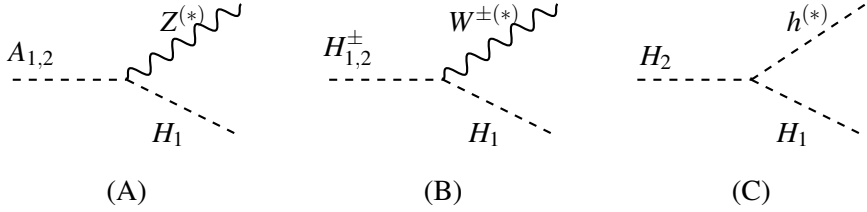
\begin{figure}[h]
\hspace{1.25cm}
\begin{tikzpicture}[thick,scale=1.0]
\draw (1.5,0) -- node[black,above,xshift=-0.1cm,yshift=0.0cm] {$ $} (1.5,0.03);
\draw[dashed] (0,0) -- node[black,above,xshift=-0.5cm,yshift=0cm] {$A_{1,2}$} (1.5,0);
\draw[decorate,decoration={snake,amplitude=3pt,segment length=10pt}] (1.5,0) -- node[black,above,xshift=0cm,yshift=0cm] {$Z^{(*)}$} (3,1.0);
\draw[dashed](1.5,0) -- node[black,above,yshift=-0.65cm,xshift=-0.2cm]  {$H_1$} (3,-0.75);
\node at (1.5,-1.5) {(A)};
\end{tikzpicture}
\hspace{0.75cm}
\begin{tikzpicture}[thick,scale=1.0]
\draw (1.5,0) -- node[black,above,xshift=-0.1cm,yshift=0.0cm] {$ $} (1.5,0.03);
\draw[dashed] (0,0) -- node[black,above,xshift=-0.5cm,yshift=0cm] {$H^\pm_{1,2}$} (1.5,0);
\draw[decorate,decoration={snake,amplitude=3pt,segment length=10pt}] (1.5,0) -- node[black,above,xshift=-0.2cm,yshift=0cm] {$W^{\pm(*)}$} (3,1.0);
\draw[dashed](1.5,0) -- node[black,above,yshift=-0.65cm,xshift=-0.2cm]  {$H_1$} (3,-0.75);
\node at (1.5,-1.5) {(B)};
\end{tikzpicture}
\hspace{0.75cm}
\begin{tikzpicture}[thick,scale=1.0]
\draw (1.5,0) -- node[black,above,xshift=-0.1cm,yshift=0.0cm] {$ $} (1.5,0.03);
\draw[dashed] (0,0) -- node[black,above,xshift=-0.5cm,yshift=0cm] {$H_{2}$} (1.5,0);
\draw[dashed] (1.5,0) -- node[black,above,xshift=0cm,yshift=0cm] {$h^{(*)}$} (3,1.0);
\draw[dashed](1.5,0) -- node[black,above,yshift=-0.65cm,xshift=-0.2cm]  {$H_1$} (3,-0.75);
\node at (1.5,-1.5) {(C)};
\end{tikzpicture}
\vspace{0.25cm}
\caption{Tree-level decays of heavy inert states into $H_1$ and on-shell or off-shell $Z$, $W^\pm$ and $h$ bosons.}
\label{tree-decays} 
\end{figure}

CP-odd and charged scalars can decay at tree-level into a lighter inert particle in association with a real(virtual) gauge boson $W^{\pm(*)}$ or $Z^{(*)}$. Assuming the mass ordering $m_{H_{1,2}} < m_{A_{1,2}} < m_{H^\pm_{1,2}}$, the  following tree-level decays appear (only diagrams with $H_1$ in the final state are shown in Fig. \ref{tree-decays}, diagrams (A) and (B)):
\be 
A_i \to Z^{(*)} H_j, \quad  H^\pm_i \to W^{\pm(*)} H_j, \quad H^\pm_i \to W^{\pm(*)} A_j, \qquad (i,j=1,2).
\ee 
The leptonic decays(splittings) of real(virtual) massive gauge bosons will result in $f \bar f$ pairs for $Z^{(*)}$ and $f \bar f'$ for $W^{\pm(*)}$.
The above processes are governed by the gauge couplings and therefore lead to small decay widths, of order $10^{-2}-10^{-4}$ GeV, of heavy inert particles. However, these decay widths could grow if the mass splitting between $H_1$ and other particles is large. Note that, even if all particle masses are relatively close (of the order of 1 GeV), they all still decay inside the detector. 

The heavy CP-even scalar, $H_2$, cannot couple to $H_1$ through $Z^{(*)}$, since CP symmetry is conserved in our model. It can decay into the $H_1$ particle plus a Higgs boson (diagram (C) in Fig. \ref{tree-decays}), which will then decay via  
the established SM patterns. Depending on the mass splitting between $H_1$ and $H_2$, the Higgs particle can be highly off-shell (recall that its SM-like nature requires its width to be around 4 MeV), thus leading to a relatively small decay width of $H_2$ and its relatively long lifetime. However, in all studied points, this width is not smaller than $10^{-11}$ GeV, ensuring the decay of $H_2$ inside the detector. 

Therefore, the $H_1$ is the only truly invisible dark particle in the benchmark scenarios we consider in the I(2+1)HDM.

\subsection{Loop-level decays of heavy inert states}
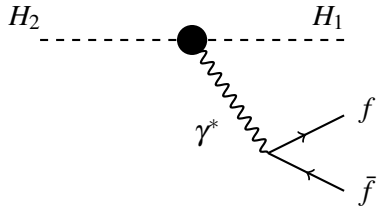
\begin{figure}[h]
           \centering
             \begin{tikzpicture}[thick,scale=1.0]
\draw[dashed] (0,0) -- node[black,above,xshift=-1.2cm,yshift=0.0cm] {$H_2$} (2,0);
\draw[dashed] (2,0) -- node[black,above,xshift=0.8cm,yshift=0.0cm] {$H_1$} (4,0);
\draw[photon] (2,0) -- node[black,above,yshift=-0.8cm,xshift=-0.3cm] {$\gamma^*$} (3,-1.5);
\draw[particle](3,-1.5) -- node[black,above,xshift=0.8cm,yshift=0.0cm] {$f$} (4,-1);
\draw[antiparticle](3,-1.5) -- node[black,above,xshift=0.8cm,yshift=-0.6cm] {$\bar f$} (4,-2);
\draw[xshift=-0cm] (2,0) node[circle,fill,inner sep=4pt](A){} -- (2,0);
\end{tikzpicture}
\caption{Radiative decay of the heavy neutral particle $H_2 \to H_1 \gamma^* \to H_1 f \bar f$.}
\label{radiative}
          \end{figure}

Apart from the above tree-level decays there is also the possibility of loop-mediated ones for a heavy neutral inert 
particle, denoted in Fig. \ref{radiative} as $H_2$, into the lightest inert state, $H_1$, and a virtual photon, which then would split into a light $f\bar f$ pair.

\begin{figure}[h]
\hspace{1.5cm}
\begin{tikzpicture}[thick,scale=1.0]
\draw (1.5,0) -- node[black,above,xshift=-0.1cm,yshift=0.0cm] {$ $} (1.5,0.03);
\draw[dashed] (0,0) -- node[black,above,xshift=-0.5cm,yshift=0cm] {$H_{2}$} (1.5,0);
\draw[decorate,decoration={snake,amplitude=3pt,segment length=10pt}] (1.5,0) -- node[black,above,xshift=0cm,yshift=0cm] {$Z$} (3,1.0);
\draw[dashed] (4.5,-0.75) -- node[black,above,yshift=-0.6cm,xshift=0cm] {$H_1$} (3,-0.75);
\draw[dashed](1.5,0) -- node[black,above,yshift=-0.65cm,xshift=-0.2cm]  {$A_{1,2}$} (3,-0.75);
\draw[particle](3,1) -- node[black,above,yshift=0.1cm,xshift=0cm]  {$f$} (4.5,1);
\draw[antiparticle](3,0.25) -- node[black,above,yshift=-0.5cm,xshift=0cm]  {$f$} (4.5,0.25);
\draw[particle](3,0.25) -- node[black,above,yshift=-0.2cm,xshift=0.3cm]  {$f$} (3,1);
\draw[decorate,decoration={snake,amplitude=3pt,segment length=10pt}] (3,-0.75) -- node[black,above,yshift=-0.3cm,xshift=0.3cm] {$Z$} (3,0.25);
\node at (1.5,-1.5) {(A)};
\end{tikzpicture}
\hspace{1.5cm}
\begin{tikzpicture}[thick,scale=1.0]
\draw (1.5,0) -- node[black,above,xshift=-0.1cm,yshift=0.0cm] {$ $} (1.5,0.03);
\draw[dashed] (0,0) -- node[black,above,xshift=-0.5cm,yshift=0cm] {$H_{2}$} (1.5,0);
\draw[decorate,decoration={snake,amplitude=3pt,segment length=10pt}] (1.5,0) -- node[black,above,xshift=0cm,yshift=0cm] {$W^+$} (3,1.0);
\draw[dashed] (4.5,-0.75) -- node[black,above,yshift=-0.6cm,xshift=0cm] {$H_1$} (3,-0.75);
\draw[dashed](1.5,0) -- node[black,above,yshift=-0.65cm,xshift=-0.2cm]  {$H^+_{1,2}$} (3,-0.75);
\draw[particle](3,1) -- node[black,above,yshift=0.1cm,xshift=0cm]  {$f$} (4.5,1);
\draw[antiparticle](3,0.25) -- node[black,above,yshift=-0.5cm,xshift=0cm]  {$f$} (4.5,0.25);
\draw[particle](3,0.25) -- node[black,above,yshift=-0.3cm,xshift=0.3cm]  {$f'$} (3,1);
\draw[decorate,decoration={snake,amplitude=3pt,segment length=10pt}] (3,-0.75) -- node[black,above,yshift=-0.4cm,xshift=0.4cm] {$W^\pm$} (3,0.25);
\node at (1.5,-1.5) {(B)};
\end{tikzpicture}\\[2mm]

\hspace{1.5cm}
\begin{tikzpicture}[thick,scale=1.0]
\draw (1.5,0) -- node[black,above,xshift=-0.1cm,yshift=0.0cm] {$ $} (1.5,0.03);
\draw[dashed] (0,0) -- node[black,above,xshift=-0.5cm,yshift=0cm] {$H_{2}$} (1.5,0);
\draw[decorate,decoration={snake,amplitude=3pt,segment length=10pt}] (1.5,0) -- node[black,above,xshift=0cm,yshift=0cm] {$Z$} (2.5,1.0);
\draw[dashed](1.5,0) -- node[black,above,yshift=-0.65cm,xshift=-0.2cm]  {$A_{1,2}$} (2.5,-1);
\draw[decorate,decoration={snake,amplitude=3pt,segment length=10pt}] (2.5,-1) -- node[black,above,yshift=-0.3cm,xshift=0.3cm] {$Z$} (3.5,0.);
\draw[particle](2.5,1) -- node[black,above,yshift=-0.6cm,xshift=0cm]  {$f$} (3.5,0);
\draw[dashed] (4.5,-1) -- node[black,above,yshift=-0.6cm,xshift=0cm] {$H_1$} (2.5,-1);
\draw[particle](3.5,0) -- node[black,above,yshift=0.0cm,xshift=0cm]  {$f$} (4.5,1);
\draw[particle](4.5,0) -- node[black,above,yshift=-0.8cm,xshift=0.7cm]  {$f$} (2.5,1);
\node at (1.5,-1.5) {(C)};
\end{tikzpicture}
\hspace{1.5cm}
\begin{tikzpicture}[thick,scale=1.0]
\draw (1.5,0) -- node[black,above,xshift=-0.1cm,yshift=0.0cm] {$ $} (1.5,0.03);
\draw[dashed] (0,0) -- node[black,above,xshift=-0.5cm,yshift=0cm] {$H_{2}$} (1.5,0);
\draw[decorate,decoration={snake,amplitude=3pt,segment length=10pt}] (1.5,0) -- node[black,above,xshift=0cm,yshift=0cm] {$W^+$} (2.5,1.0);
\draw[dashed](1.5,0) -- node[black,above,yshift=-0.65cm,xshift=-0.2cm]  {$H^+_{1,2}$} (2.5,-1);
\draw[decorate,decoration={snake,amplitude=3pt,segment length=10pt}] (2.5,-1) -- node[black,above,yshift=-0.3cm,xshift=0.5cm] {$W^+$} (3.5,0.);
\draw[particle](2.5,1) -- node[black,above,yshift=-0.6cm,xshift=0cm]  {$f'$} (3.5,0);
\draw[dashed] (4.5,-1) -- node[black,above,yshift=-0.6cm,xshift=0cm] {$H_1$} (2.5,-1);
\draw[particle](3.5,0) -- node[black,above,yshift=0.0cm,xshift=0cm]  {$f$} (4.5,1);
\draw[particle](4.5,0) -- node[black,above,yshift=-0.8cm,xshift=0.7cm]  {$f$} (2.5,1);
\node at (1.5,-1.5) {(D)};
\end{tikzpicture}
\caption{Box diagrams contributing to $H_2 \to H_1 f \bar f$. }
\label{box} 
\end{figure}
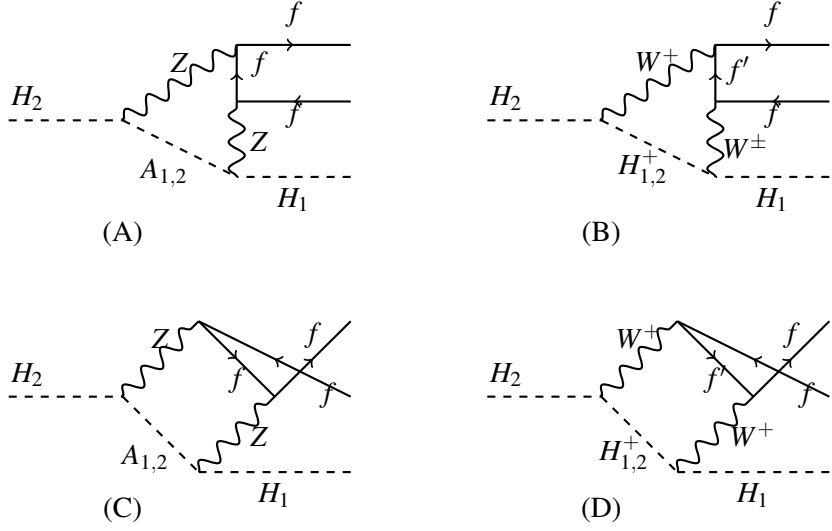

The corresponding loops go through triangle and bubble diagrams with $H^\pm_i$ and $W^\pm$ entering, see Figs   
 \ref{triangle-decays}-\ref{bubble-decays}. 
Note that there are also box diagrams which contribute to the process $H_2 \to H_1 f \bar f$, presented in Fig. \ref{box}. Here, the $f \bar f$ pair is produced through the SM gauge-fermion tree-level vertices, without producing an intermediate off-shell photon. The corresponding topologies also see the contribution of inert, both charged and neutral (pseudo)scalars.
However, due to the mass suppression, the contribution from the box diagrams is  small, of order 10\%, and it leaves the results practically unaffected. 
For reasons of optimisation then,  we do not show the results  of these box diagrams in the numerical scans and we may refer to this one-loop process as a radiative decay. 

Note that the process $A_i\to H_1 Z^*$ does exist at tree-level in both the I(2+1)HDM (for $i=1,2$)  and  I(1+1)HDM (for $i=1$) and contributes to the $\Et  f \bar f$ signature, as discussed previously. However, in the interesting regions of the parameter space where the invariant mass of the $f \bar f$ pair is small, i.e., $<<m_Z$, this process is sub-dominant.

In short, the only (effective) loop-level decay to consider is 
\be
H_2 \to H_1 \gamma^*
\ee
and this does not exist in the I(1+1)HDM, as CP-conservation prevents the only possibly similar  radiative decay in its inert sector (i.e., $A_1\to H_1\gamma^*$). Therefore, as intimated, this signature can be used to distinguish between the  I(1+1)HDM and models with extended inert sectors, such as the I(2+1)HDM.

\begin{figure}[h]
\hspace{1.5cm}
\begin{tikzpicture}[thick,scale=1.0]
\draw (1.5,0) -- node[black,above,xshift=-0.1cm,yshift=0.0cm] {$ $} (1.5,0.03);
\draw[dashed] (0,0) -- node[black,above,xshift=-0.5cm,yshift=0cm] {$H_{2}$} (1.5,0);
\draw[dashed] (1.5,0) -- node[black,above,xshift=0cm,yshift=0cm] {$H^+_{1,2}$} (3,1.0);
\draw[dashed] (4.5,-0.75) -- node[black,above,yshift=-0.1cm,xshift=0cm] {$H_1$} (3,-0.75);
\draw[decorate,decoration={snake,amplitude=3pt,segment length=10pt}](1.5,0) -- node[black,above,yshift=-0.65cm,xshift=-0.2cm]  {$W^+$} (3,-0.75);
\draw[dashed](3,1) -- node[black,above,yshift=-0.4cm,xshift=-0.4cm]  {$H^+_{1,2}$} (3,-0.75);
\draw[decorate,decoration={snake,amplitude=3pt,segment length=10pt}](3,1) -- node[black,above,yshift=0.1cm,xshift=0cm]  {$\gamma^*$} (4.5,1);
\node at (1.5,-1.5) {(A)};
\end{tikzpicture}
\hspace{1.5cm}
\begin{tikzpicture}[thick,scale=1.0]
\draw (1.5,0) -- node[black,above,xshift=-0.1cm,yshift=0.0cm] {$ $} (1.5,0.03);
\draw[dashed] (0,0) -- node[black,above,xshift=-0.5cm,yshift=0cm] {$H_{2}$} (1.5,0);
\draw[decorate,decoration={snake,amplitude=3pt,segment length=10pt}] (1.5,0) -- node[black,above,xshift=0cm,yshift=0cm] {$W^+$} (3,1.0);
\draw[dashed] (4.5,-0.75) -- node[black,above,yshift=-0.1cm,xshift=0cm] {$H_1$} (3,-0.75);
\draw[dashed](1.5,0) -- node[black,above,yshift=-0.65cm,xshift=-0.2cm]  {$H^+_{1,2}$} (3,-0.75);
\draw[decorate,decoration={snake,amplitude=3pt,segment length=10pt}](3,1) -- node[black,above,yshift=-0.4cm,xshift=-0.4cm]  {$W^+$} (3,-0.75);
\draw[decorate,decoration={snake,amplitude=3pt,segment length=10pt}](3,1) -- node[black,above,yshift=0.1cm,xshift=0cm]  {$\gamma^*$} (4.5,1);
\node at (1.5,-1.5) {(B)};
\end{tikzpicture}
\vspace{0.5cm}
\caption{Triangle diagrams contributing to the $H_2 \to H_1 \gamma^*$ decay, where the lightest inert is absolutely stable and hence invisible,
while $\gamma^*$ is a virtual photon that couples to fermion-antifermion pairs.  Analogous diagrams cannot be constructed if the initial particle is $A_{1}$ or $A_2$.}
\label{triangle-decays} 
\end{figure}
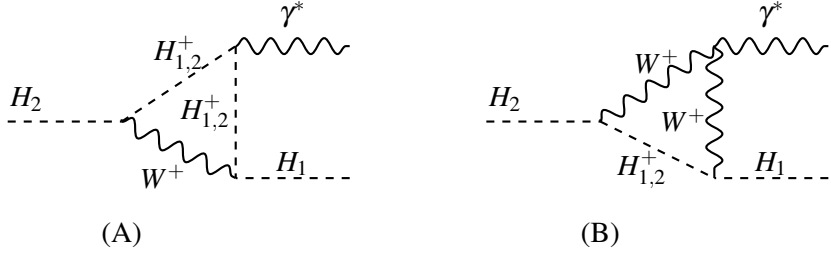    

\begin{figure}[h]
\begin{tikzpicture}[thick,scale=1.0]
\draw[dashed] (0,0) -- node[black,above,sloped,yshift=-0.1cm,xshift=-0.4cm] {$H_{2}$} (1,0);
\draw[dashed] (3,0) -- node[black,above,yshift=-0.3cm,xshift=0.5cm] {$H_1$} (4.2,-1.1);
\draw[decorate,decoration={snake,amplitude=3pt,segment length=10pt}] (3,0) -- node[black,above,yshift=-0.4cm,xshift=0.5cm] {$\gamma^*$} (4.2,1.1);
\draw[dashed]  (1,0) node[black,above,sloped,yshift=0.95cm,xshift=1.05cm] {$H^+_{1,2}$}  arc (180:0:1cm) ;
\draw[decorate,decoration={snake,amplitude=3pt,segment length=10pt}]  (1,0) node[black,above,sloped,yshift=-0.95cm,xshift=1.05cm] {$W^+$}  arc (-180:0:1cm) ;
\node at (1.5,-2.3) {(A)};
\end{tikzpicture}
\hspace{2mm}
\begin{tikzpicture}[thick,scale=1.0]
\draw[dashed] (0,0) -- node[black,above,xshift=-0.3cm,yshift=-0.1cm] {$H_{2}$} (1.5,0);
\draw[dashed] (1.5,0) -- node[black,above,yshift=0.4cm,xshift=0.2cm] {$H_1$} (2.7,1.4);
\draw[decorate,decoration={snake,amplitude=3pt,segment length=10pt}] (2.8,-1.2) -- node[black,above,yshift=-0.3cm,xshift=0.9cm] {$\gamma^*$} (4.3,-1.7);
\draw[dashed]  (1.5,0) node[black,above,sloped,yshift=0.2cm,xshift=1.3cm] {$H^+_{1,2}$}  arc (140:-40:0.9cm) ;
\draw[dashed]  (1.5,0) node[black,above,sloped,yshift=-1.4cm,xshift=0.55cm] {$H^+_{1,2}$}  arc (-220:-40:0.9cm) ;
\node at (1.5,-2.3) {(B)};
\end{tikzpicture}
\begin{tikzpicture}[thick,scale=1.0]
\draw[dashed] (0,0) -- node[black,above,xshift=-0.3cm,yshift=-0.1cm] {$H_{2}$} (1.5,0);
\draw[decorate,decoration={snake,amplitude=3pt,segment length=10pt}] (1.5,0) -- node[black,above,yshift=0.4cm,xshift=0.2cm] {$\gamma^*$} (2.7,1.4);
\draw[dashed] (2.8,-1.2) -- node[black,above,yshift=-0.3cm,xshift=0.9cm] {$H_1$} (4.3,-1.7);
\draw[dashed]  (1.5,0) node[black,above,sloped,yshift=0.2cm,xshift=1.3cm] {$H^+_{1,2}$}  arc (140:-40:0.9cm) ;
\draw[decorate,decoration={snake,amplitude=3pt,segment length=10pt}]  (1.5,0) node[black,above,sloped,yshift=-1.4cm,xshift=0.55cm] {$W^+$}  arc (-220:-40:0.9cm) ;
\node at (1.5,-2.3) {(C)};
\end{tikzpicture}
\vspace{0.5cm}
\caption{Bubble  diagrams contributing to the $H_2 \to H_1 \gamma^*$ decay, where the lightest inert particle is absolutely stable and hence invisible,
while $\gamma^*$ is a virtual photon that couples to fermion-antifermion pairs.  Analogous diagrams cannot be constructed if the initial particle is $A_{1}$ or $A_2$.}
\label{bubble-decays} 
\end{figure}
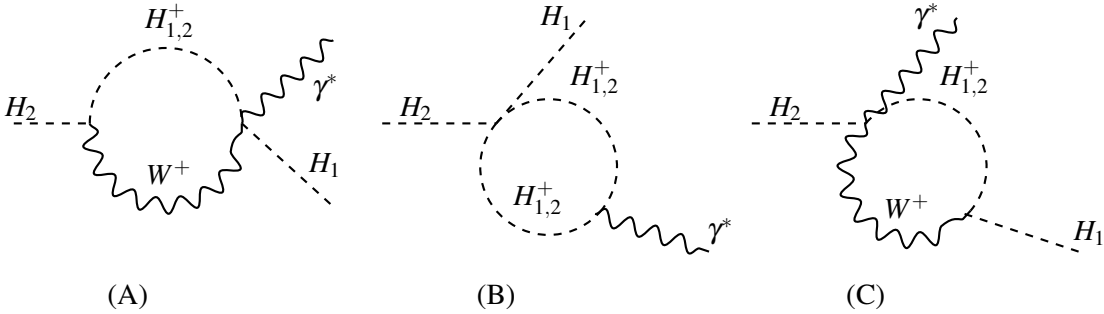    

\subsection{The $\Et$ $f \bar f$ signature at the LHC}

In this subsection, we focus on the possible sources of the aforementioned specific signature that can arise in the I(2+1)HDM, namely,  missing transverse energy and a fermion-antifermion pair, $\Et f \bar f$. This final state can be produced both at tree-level and through one-loop decays, as previously explained. We dwell further on this here.

\begin{figure}[h]
           \centering
             \begin{tikzpicture}[thick,scale=1.0]
\draw[gluon] (0,0) -- node[black,above,xshift=-0.6cm,yshift=0.4cm] {$g$} (1,-1);
\draw[gluon] (0,-2) -- node[black,above,yshift=-1.0cm,xshift=-0.6cm] {$g$} (1,-1);
\draw[dashed] (1,-1) -- node[black,above,xshift=0.0cm,yshift=0.0cm] {$h$} (2.5,-1);
\draw[dashed] (2.5,-1) -- node[black,above,xshift=0.6cm,yshift=0.4cm] {$S_i$} (3.5,0);
\draw[dashed] (2.5,-1) -- node[black,above,yshift=-1cm,xshift=0.4cm] {$S_j$} (3.5,-2);
\draw[xshift=-0cm] (1,-1) node[circle,fill,inner sep=4pt](A){} -- (1,-1);
\end{tikzpicture}
\caption{The ggF-induced production of the SM-like Higgs particle at the LHC with its decay into inert particles, denoted as $S_i$ and $S_j$.}
\label{Higgsprod}
          \end{figure}
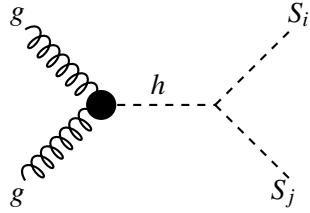

The first mechanism is related to decays of the SM-like Higgs particle which is produced, e.g., through ggF.
The $hgg$ effective vertex is identical to that in the SM, as the gauge and fermionic sectors in the I(2+1)HDM are not modified with respect to the SM. The Higgs particle can then decay into a pair of neutral or charged inert particles, denoted in Fig. \ref{Higgsprod} by $S_{i,j}$. Depending on the masses of $S_{i,j}$, these particles can further decay, providing various final states.

In the CPC I(2+1)HDM, a process contributing to the $\Et f \bar f$ signature (and one of our signals) is 
\be
gg \to h \to H_1 H_2 \to H_1 H_1 \gamma^* \to H_1 H_1 f \bar f,
\label{first}
\ee
where the off-shell $\gamma^*$ splits into $f \bar f$ and the $H_1$ states escape detection. 

Notice that there is also a tree-level $h$ decay into two charged scalars with the same signature ($\Et\;  f \bar f$), albeit not an identical final state (the two would remain indistinguishable though), following the pattern:
\be
gg \to h \to H^{\pm}_i H^{\pm}_i \to  H_1  H_1 W^{+(*)} W^{-(*)}\to H_1 H_1 \nu_l l^+ \nu_l l^-  \quad (i=1,2),
\label{second}
\ee 
where the neutrinos escape detection as (additional) $\Et$.

The  process in  (\ref{first}) is loop-mediated and depends on $g_{H_1 H_2 h}$, a coupling affecting also DM relic density. Therefore, if this coupling is small, the whole process is  suppressed. However, we shall maximize this coupling,
while maintaining consistency with DM constraints. We also assume a mass spectrum so that the charged Higgs masses
entering the loops are not too heavy, since their large masses would also suppress the loop. 
In fact, we shall see that there can be parameter configurations for which $m_{H_1}+m_{H_2}<m_h$, so that 
SM-like Higgs production and (loop) decay is resonant, thereby benefiting of an enhancement of ${\cal O}(1/\alpha_{\rm EM})$.
The  process in  (\ref{second}) is a tree-level one, therefore potentially competitive. However, 
for the parameter space of interest, maximizing the yield of the loop process, this mode becomes negligible, for two reasons: on the one hand, the charged Higgs masses are generally heavy so that there can be no resonant $h$ involved while, on the other hand, the  $g_{H^\pm_i H^\pm_ih}$ coupling is generally small. 

\begin{figure}[h]
\begin{tikzpicture}[thick,scale=1.0]

\hspace{-1cm}

\draw[particle] (0,0) -- node[black,above,xshift=-0.6cm,yshift=0.4cm] {$q_i$} (1,-0.75);
\draw[antiparticle] (0,-1.5) -- node[black,above,yshift=-1.0cm,xshift=-0.6cm] {$\bar{q}_i$} (1,-0.75);
\draw[photon] (1,-0.75) -- node[black,above,xshift=0.0cm,yshift=0.0cm] {$Z^*$} (2,-0.75);
\draw[dashed] (2,-0.75) -- node[black,above,yshift=0.0cm,xshift=-0.0cm] {$h$} (3,-0.75);
\draw[dashed] (3,-0.75) -- node[black,above,yshift=0.1cm,xshift=0.8cm] {$H_1$} (4,-0);
\draw[dashed] (3,-0.75) -- node[black,above,yshift=-0.4cm,xshift=0.8cm] {$H_1$} (4,-1.5);
\draw[photon] (2,-0.75) -- node[black,above,xshift=1.1cm,yshift=-1.5cm] {$Z$} (4,-2.5);

\node at (2,-4) {(A)};

\hspace{1cm}

\draw[particle] (5,0) -- node[black,above,xshift=-0.6cm,yshift=0.4cm] {$q$} (6,-0.75);
\draw[antiparticle] (5,-1.5) -- node[black,above,yshift=-1.0cm,xshift=-0.6cm] {$\bar{q}$} (6,-0.75);
\draw[photon] (6,-0.75) -- node[black,above,xshift=0.0cm,yshift=0.0cm] {$Z^*$} (7,-0.75);
\draw[dashed] (7,-0.75) -- node[black,above,yshift=0.1cm,xshift=1.2cm] {$H_1$} (8.5,-0);
\draw[dashed] (7,-0.75) -- node[black,above,yshift=-0.4cm,xshift=1.1cm] {$H_1$} (8.5,-1.5);
\draw[photon] (7,-0.75) -- node[black,above,xshift=1.15cm,yshift=-1.5cm] {$Z$} (8.5,-2.5);

\node at (6,-4) {(B)};

\hspace{1cm}

\draw[particle] (10,0) -- node[black,above,xshift=-0.6cm,yshift=0.4cm] {$q$} (11,-0.75);
\draw[antiparticle] (10,-1.5) -- node[black,above,yshift=-1.0cm,xshift=-0.6cm] {$\bar{q}$} (11,-0.75);
\draw[photon] (11,-0.75) -- node[black,above,xshift=0.0cm,yshift=0.0cm] {$Z^*$} (12,-0.75);
\draw[dashed] (12,-0.75) -- node[black,above,yshift=0.1cm,xshift=0.8cm] {$H_1$} (13,-0);
\draw[dashed] (12,-0.75) -- node[black,above,yshift=-0.3cm,xshift=-0.3cm] {$A_{1,2}$} (12,-1.75);
\draw[dashed] (12,-1.75) -- node[black,above,yshift=0.1cm,xshift=0.8cm] {$H_1$} (13,-1);
\draw[photon] (12,-1.75) -- node[black,above,xshift=0.6cm,yshift=-0.9cm] {$Z^{(*)}$} (13,-2.5);
\node at (11,-4) {(C)};
\end{tikzpicture}
\caption{Diagrams leading to the $\Et f \bar f$ final state via the $H_1H_1 Z^{(*)}$ intermediate stage.}
\label{diag:nstr}
\vspace*{0.75cm}
\end{figure}
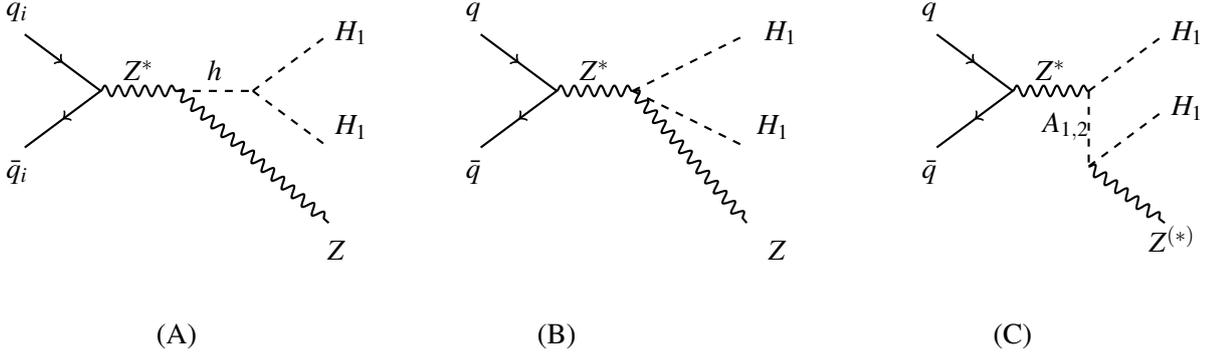

In principle, there is another tree-level signal inducing the $\Et\; f \bar f$ final state in our scenario,
\begin{equation}
\label{nstr}
q\bar q \to Z^*\to H_1H_1 Z\to H_1H_1 f \bar f,
\end{equation}
see diagrams  (A) and (B) in Fig. \ref{diag:nstr},  induced by quark-antiquark annihilation and proceeding via an $s$-channel off-shell (primary) $Z^*$, wherein the on-shell (secondary) $Z$ eventually decays into a $f \bar f$ pair. However, this is of no concern here. The reason is twofold. On the one hand, as explained, the region of parameter space over which process (\ref{first}) is interesting for LHC phenomenology is the one where the $g_{H_1 H_2 h}$ strength is maximal and $h$ is possibly resonant: this is when the DM relic density sees a large contribution from $H_1H_2$ co-annihilation processes\footnote{This is further enhanced when $m_{H_1}\approx m_{H_2}$, which is in fact one of the conditions that we will use in the forthcoming analysis to exalt   process (\ref{first}) (which is I(2+1)HDM specific) against  the one (also existing in the I(1+1)HDM) that we will be discussing next.}, which in turn means that large $g_{H_1 H_1 h}$ (possibly in presence of a resonant $h$) and $g_{H_1H_1 ZZ}$  couplings are forbidden by such  data, so that process (\ref{nstr}) becomes uninteresting at the LHC. On the other hand, in our construct,  process (\ref{nstr}) is nothing more than a subleading contribution to the invisible Higgs signature of the SM-like Higgs boson (dominated by ggF and VBF topologies, extensively studied already in Ref.~\cite{Keus:2014isa}), rather featureless, in fact, as it does not catch any of the heavy scalar states of the model, 
unlike reaction (\ref{first}), which is sensitive to all of them, so that one could study the kinematic distributions of the 
final state attempting to extract their masses by isolating the corresponding thresholds entering the loops\footnote{In this sense, process (\ref{nstr}) would be
a background to (\ref{first}), which can be easily removed through a mass veto: $m_{f \bar f}\ne m_Z$.}. For these reasons, we will not discuss these two topologies any further. 

Another way of obtaining exactly the $H_1 H_1  f \bar f$ final state is shown in graph (C) of Fig. \ref{diag:nstr}, again produced through $s$-channel quark-antiquark annihilation into a virtual neutral massive  gauge boson, i.e., 
\begin{equation}
\label{tree}
q\bar q\to Z^* \to H_1 A_i \to H_1 H_1 Z^{(*)}  \to H_1 H_1 f \bar f\qquad(i=1,2),
\end{equation}
wherein the DM candidate is produced in association with a pseudoscalar state and the $Z$ may be off-shell.
This mode is indeed competitive with the one in  (\ref{first}) over the region of I(2+1)HDM parameter space of
interest, so we will extensively dwell with it numerically in the remainder of the paper. Further, diagram (C) in Fig. \ref{diag:nstr}, unlike graphs (A) and (B) herein, because of its heavy pseudoscalar components, may also be isolated in 
the aforementioned kinematic analysis.

\begin{figure}[h]
             \begin{tikzpicture}[thick,scale=1.0]
\draw[particle] (0,0) -- node[black,above,xshift=-0.6cm,yshift=0.0cm] {$q_i$} (1,0);
\draw[photon] (1,0) -- node[black,above,xshift=-0.8cm,yshift=-0.3cm] {$Z,W^+$} (1,-1.5);
\draw[photon] (1,-1.5) -- node[black,above,xshift=-0.8cm,yshift=-0.2cm] {$Z,W^+$} (1,-3);
\draw[particle] (0,-3) -- node[black,above,xshift=-0.6cm,yshift=-0.6cm] {$q_j$} (1,-3);
\draw[dashed] (1,-1.5) -- node[black,above,yshift=0.1cm,xshift=-0.0cm] {$h$} (2.5,-1.5);
\draw[dashed] (2.5,-1.5) -- node[black,above,yshift=0.3cm,xshift=0.6cm] {$H_1$} (3.5,-0.75);
\draw[dashed] (2.5,-1.5) -- node[black,above,yshift=-1.1cm,xshift=0.6cm] {$H_2$} (3.5,-2.25);
\draw[particle] (1,0) -- node[black,above,xshift=1.2cm,yshift=0.0cm] {$q_k$} (3.5,0);
\draw[particle] (1,-3) -- node[black,above,xshift=1.2cm,yshift=-0.6cm] {$q_l$} (3.5,-3);

\node at (2,-4) {(A)};

\hspace{1cm}

\draw[particle] (5,0) -- node[black,above,xshift=-0.6cm,yshift=0.0cm] {$q_i$} (6,0);
\draw[photon] (6,0) -- node[black,above,xshift=-0.8cm,yshift=-0.3cm] {$Z,W^+$} (6,-1.5);
\draw[photon] (6,-1.5) -- node[black,above,xshift=-0.8cm,yshift=-0.2cm] {$Z,W^+$} (6,-3);
\draw[particle] (5,-3) -- node[black,above,xshift=-0.6cm,yshift=-0.6cm] {$q_j$} (6,-3);
\draw[particle] (6,0) -- node[black,above,xshift=0.8cm,yshift=0.0cm] {$q_k$} (7.5,0);
\draw[particle] (6,-3) -- node[black,above,xshift=0.8cm,yshift=-0.6cm] {$q_l$} (7.5,-3);
\draw[dashed] (6,-1.5) -- node[black,above,yshift=0.3cm,xshift=0.6cm] {$H_1$} (7.5,-0.75);
\draw[dashed] (6,-1.5) -- node[black,above,yshift=-1.1cm,xshift=0.6cm] {$H_2$} (7.5,-2.25);

\node at (6,-4) {(B)};

\hspace{1cm}

\draw[particle] (8.5,0) -- node[black,above,xshift=-0.6cm,yshift=0.0cm] {$q_i$} (9.5,0);
\draw[photon] (9.5,0) -- node[black,above,xshift=-0.9cm,yshift=-0.3cm] {$Z (W^+)$} (9.5,-1);
\draw[dashed] (9.5,-1) -- node[black,above,xshift=-0.9cm,yshift=-0.3cm] {$A_1 (H^+_1)$} (9.5,-2);
\draw[photon] (9.5,-2) -- node[black,above,xshift=-0.9cm,yshift=-0.2cm] {$Z (W^+)$} (9.5,-3);
\draw[particle] (8.5,-3) -- node[black,above,xshift=-0.6cm,yshift=-0.6cm] {$q_j$} (9.5,-3);
\draw[particle] (9.5,0) -- node[black,above,xshift=0.8cm,yshift=0.0cm] {$q_k$} (11,0);
\draw[particle] (9.5,-3) -- node[black,above,xshift=0.8cm,yshift=-0.6cm] {$q_l$} (11,-3);
\draw[dashed] (9.5,-1) -- node[black,above,yshift=0.1cm,xshift=0.6cm] {$H_1$} (11,-1);
\draw[dashed] (9.5,-2) -- node[black,above,yshift=-0.8cm,xshift=0.6cm] {$H_2$} (11,-2);

\node at (10,-4) {(C)};

\end{tikzpicture}
              \caption{Diagrams  leading to the $\Et + f \bar f$ final state via VBF topologies.}
\label{diag:VBF}
          \end{figure}
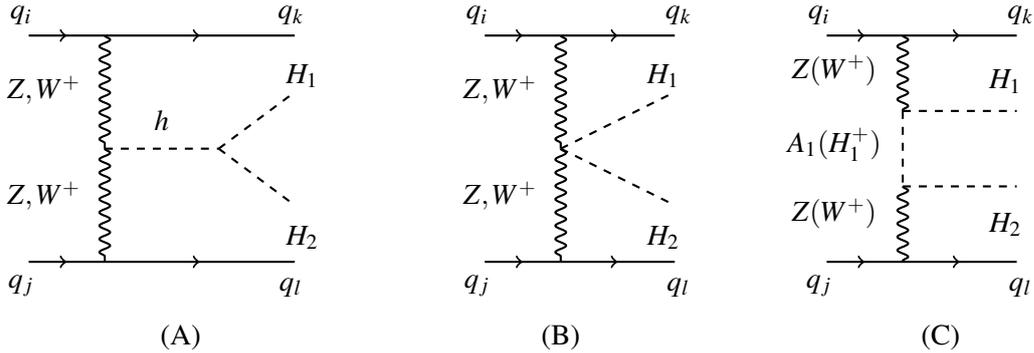

Finally, we conclude this subsection by listing, in Fig. \ref{diag:VBF} (prior to the $H_2\to H_1 f \bar f$ decay), the topologies entering VBF production contributing to
the $\Et\; f \bar f$ final state (our second signal) via
\begin{equation}
\label{VBF}
q_i q_j \to q_k q_l H_1 H_2 \to H_1 H_1 \gamma^* \to H_1 H_1 f \bar f,
\end{equation}
where $q_{i,j,k,l}$ represents a(n) (anti)quark of any possible flavour (except a top quark).
Here, two aspects are worth noticing. Firstly, there is the additional presence  of two forward/backward jets, which may or may not be tagged (we will treat them inclusively). Secondly, not all diagrams 
proceed via $h\to H_1H_2$ induced topologies, graph (A), hence unlike the case of ggF, since graphs (B) and (C)
are also possible. Clearly, the first diagram dominates when $h$ can resonate while the last two become competitive otherwise. 
We shall see how ggF and VBF will compete over the I(2+1)HDM parameter space of interest in being the carrier of its hallmark signature
$\Et \; f \bar f$ in a later section.

\section{Calculation \label{sec-loop}}

In this section, we discuss the details of our  calculation. In fact, the case of the channel in  (\ref{tree}) is easily dealt with, as this
is a tree-level process, which we computed numerically using CalcHEP \cite{Belyaev:2012qa}. 
The bulk of our effort was  concentrated upon the loop processes (\ref{first}) and (\ref{VBF}),
which we have tackled in factorised form, i.e., by breaking up the two channels into $pp\to H_1 H_2 X$ production followed
by the $H_2$ $\to$ $H_1  f \bar f$ decay. Here, the ggF and VBF topologies entering at production level are well known in the literature, so we do not discuss them (again, we computed these numerically by exploiting CalcHEP). We therefore address in some detail only the case of the loop decay.

This is expressed through a tensor structure appropriate to the I(2+1)HDM particle spectrum and illustrated for the case $f=e$, so that we can safely take $m_e=0$\footnote{The case $f=u,d,c,s,\mu,\tau$ with $m_f\ne0$ is a straightfoward extension of it.}. 
In general, there are two types of one-loop diagrams that contribute to the process 
$$H_2(p_3) \to H_1(p_2) \gamma^* (p_3-p_2) \to H_1(p_2) e^- (k_1) e^+ (k_2),$$
namely, those embedding the one-loop effective vertex $H_2 H_1 \gamma^*$, given by the diagrams in  Figs. 
\ref{triangle-decays}--\ref{bubble-decays} plus the box diagrams shown in Fig. \ref{box}. Here, the labels $p_i$ and $k_j$ identify the external scalar and fermion momenta, respectively. In the following, we use the unitary gauge.

The general expression for the amplitude of the loop calculation is:
\begin{equation}
 \mathcal{M}=ie\bar{v}(k_1)\gamma^\nu u(k_2)\frac{ig_{\mu\nu}}{(p_3-p_2)^2}[A(p_3+p_2)^\mu +B(p_3-p_2)^\mu],
\end{equation}
where
\begin{equation}
i[A(p_3+p_2)^\mu +B(p_3-p_2)^\mu]
\end{equation}
is the general structure of the vertex $H_1  H_2 \gamma^*$ obtained in the calculation at one-loop level.
However, when we consider the term $(p_3-p_2)^\nu$ and contract it with $\gamma^\mu g_{\mu\nu}$ we have:
 \begin{eqnarray}
\slashed{p}_3-\slashed{p}_2=\slashed{k}_1-\slashed{k}_2. 
\end{eqnarray}
Then the Dirac equation in the limit of $m_e=0$ gives us:
\begin{eqnarray}
\bar{v}(k_1)(\slashed{p}_3-\slashed{p}_2)u(k_2)=\bar{v}(k_1)(\slashed{k}_1+\slashed{k}_2)u(k_2)=0. 
\end{eqnarray}
Under these circumstances,  we can take 
 \[(p_3-p_2)_\mu =0,  \]
which is the same as if the $\gamma$ were on-shell in the process $H_2\to H_1\gamma$, albeit  $(p_3-p_2)^2$ is non-zero:
\begin{equation}
 (p_3-p_2)^2=(k_1+k_2)^2=2k_1\cdot k_2.
\end{equation}
Therefore, the general structure of the amplitude is:
\begin{eqnarray}
 \mathcal{M}=ie\bar{v}(k_1)\gamma^\nu u(k_2)\frac{ig_{\mu\nu}}{(p_3-p_2)^2}[A(p_3+p_2)^\mu ],
\label{Amp1}
\end{eqnarray} 
where $A(p_3+p_2)^\mu$ is related to the contribution of the each diagram in Figs. \ref{triangle-decays}, \ref{bubble-decays} and \ref{box}:
\begin{eqnarray}
A(p_3+p_2)^\mu =M_{\mu,T}= \sum_i M_\mu^{(i)},
\label{Amp2}
\end{eqnarray}
where $i$ runs across all diagrams. 

Then the square amplitude (\ref{Amp1}) of the loop process is (upon the usual final state spin summation):
\begin{eqnarray}
|\mathcal{M}|^2 &= &\frac{8 |A|^2}{m_{12}^4} \bigg((m_{H_2}^2- m_{23}^2 ) (m_{23}^2 - m_{H_1}^2)   -m_{12}^2 m_{23}^2\bigg), 
 \label{Ampsq1}
\end{eqnarray}  
where $m^2_{12} = (p_3 - p_2)^2 = (k_1 + k_2)^2 = 2k_1\cdot k_2$ and $m^2_{23} = (k_2 - p_2)^2 = 2k_2\cdot p_2 + m^2_{H_1}$.

In agreement with Ref. \cite{Agashe:2014kda}, the partial decay width of  $H_2 \to H_1 e^- e^+$ is:
\begin{eqnarray}
\Gamma= \frac{1}{256 \pi^3 m_{H_2}^3} \int_{0}^{(m_{H_2}-m_{H_1})^2}dm_{12}^2  \Bigg( \int_{(m_{23}^2)_{min}}^{(m_{23}^2)_{max}}  d m_{23}^2 |\mathcal{M}|^2 \Bigg). \label{wh2}
\end{eqnarray}

Now, in order to perform our calculations of the cross-section using CalcHEP, we can implement the effective $H_2H_1e^+e^-$ vertex in LanHEP\cite{Semenov:2014rea}/CalcHEP. We can induce the effective vertex as:
\begin{eqnarray}
L_{(H_2 H_1 e^+ e^-)}
 &=& i K (H_1 \partial_\mu H_2- H_2 \partial_\mu H_1) \bar{e}  \gamma^\mu e.
\end{eqnarray}

If we now calculate the amplitude from this effective Lagrangian we see that
\begin{eqnarray}
M=i K \bar{v}(k_1) \gamma^\mu (p_3+p_2)_{\mu} u(k_2) 
\end{eqnarray}
so that the amplitude squared is
\begin{eqnarray}
|M|^2= 8 |K|^2  \lambda(m_{H_2}, m_{H_1},m_{23}^2).
\end{eqnarray}
Thus, the partial decay rate of the $H_2\to H_1 e^- e^+$ channel, in terms of a $K$-factor, is
\begin{eqnarray}
\Gamma&=& \frac{1}{256 \pi^3 m_{H_2}^3} \int_{0}^{(m_{H_2}-m_{H_1})^2}dm_{12}^2  \Bigg( \int_{(m_{23}^2)_{min}}^{(m_{23}^2)_{max}}  d m_{23}^2 |M|^2 \Bigg) \\
&=&\frac{1}{16 \pi^3 m_{H_2}^3}   |K|^2 \int_{0}^{(m_{H_2}-m_{H_1})^2} d m_{12}^2   I_2 = \frac{1}{16 \pi^3 m_{H_2}^3}   |K|^2 I_3, 
\end{eqnarray}
where $I_3$ is given by 
\begin{eqnarray}
I_3 & =& \int_{0}^{(m_{H_2}-m_{H_1})^2} d m_{12}^2   I_2. 
\end{eqnarray}
Finally, the $K$-factor is therefore  given by   
\begin{eqnarray}
K^2= \frac{16 \pi^3 m_{H_2}^3 \Gamma(H_2 \to H_1 e^+ e^-)}{I_3},
\label{K-factor}
\end{eqnarray}
where the width $\Gamma(H_2 \to H_1 e^+ e^-)$ is  calculated using LoopTools \cite{Hahn:1998yk}. 
Thus, the $K$-factor is related directly to the loop calculation through (\ref{K-factor}). 
Using this method, we are able to use CalcHEP, since we no longer need to 
perform any integration externally to the generator itself,  as required by the fully fledged computation performed in the previous subsection, thereby by-passing the fact that CalcHEP is actually a tree-level generator. 


\section{Results}
\label{results}

The benchmark scenarios that we study here do not necessarily correspond to regions of the parameter space where our DM candidate accounts for all the observed relic density in agreement with Planck data. In fact, the aim of these benchmark scenarios is to show in which regions of the parameter space the model has a discovery potential at the LHC.
In here, we define 
three base benchmark scenarios, A50, I5 and I10 in the low DM mass region ($m_{H_1} \leq 90$ GeV) as shown in Table 1. 

The main distinguishing parameter for the present calculation is the mass splitting between $H_1$ and the other CP-even scalar, $H_2$. 
Benchmark A50 ($m_{H_2}-m_{H_1}=50$ GeV) is taken from the analysis done in \cite{Keus:2014jha}. Relatively large mass splittings between $H_1$ and other neutral scalars leads to a standard DM annihilation in the Universe, providing us with a DM candidate which is in agreement with DM searches for a large part of the parameter space. However, we expect the tree-level decays to dominate over the loop signal through the $H_1 A_1 Z$ vertex. 

Benchmarks I5 ($m_{H_2}-m_{H_1}=5$ GeV) and I10 ($m_{H_2}-m_{H_1}=10$ GeV) have an intermediate-mass splitting between $H_1$ and $H_2$ of the order of a few GeV. For these benchmarks, we expect the tree-level decays to be reduced since there is a small mass gap between $H_1$ and $A_{1,2}$. Therefore, the intermediate gauge boson is produced off-shell. 
Further decreasing of the $H_1$-$H_2$ and $H_1$-$A_{1,2}$ mass splittings\footnote{One needs to take extra care with very small mass splittings, as they might lead to a large particle lifetime which will cause the particle to decay outside the detector.}, leads to a strengthening of the desired loop signal, with the further reduction of all tree-level decays. Note, however, that with increasing mass splitting, the loop process acquires more phase space and starts seeing the $Z^* \to ll$ contribution and the partial width grows as a result.

In all cases, differences between $m_{H_1}$ and masses of both charged scalars are relatively large. This leads to important consequences for the thermal history of DM particles: charged scalars are short-lived, and they will not take part in the freeze-out process of $H_1$. However, this mass difference is not big enough to suppress the studied loop processes. 
Increasing this mass difference would lead to a smaller cross-section and, therefore,  worse detection prospects. We would also like to stress that the all chosen mass splittings are in agreement with Electro-Weak Precision Data (EWPD) constraints, which disfavour a significant discrepancy between masses of charged and neutral particles. On the other hand, a significant reduction of this mass splitting would increase the co-annihilation effect in the Universe, hence leading to heavily reduced relic density, and thus disfavouring the I(2+1)HDM as the model for DM.

\begin{table}[h!]
	\centering
\label{BPs}
\begin{tabular}{|c||c|c|c|c|c|}
\hline
Benchmark & $m_{H_2} - m_{H_1}$ & $m_{A_1} - m_{H_1}$ & $m_{A_2} - m_{H_1}$  & $m_{H^\pm_1} - m_{H_1}$ & $m_{H^\pm_2} - m_{H_1}$ \\
\hline
A50 & 50 & 75 & 125 & 75 & 125  \\
\hline
I5 &	5 & 10	& 15 & 90 & 95  \\
\hline
I10 &	 10  & 20 &30	&  90 & 100  \\
\hline
\end{tabular}
\caption{Definition of benchmark scenarios with the mass splittings shown in GeV.}
\end{table}

\begin{figure}[h!]
\begin{center}
\includegraphics[scale=0.3]{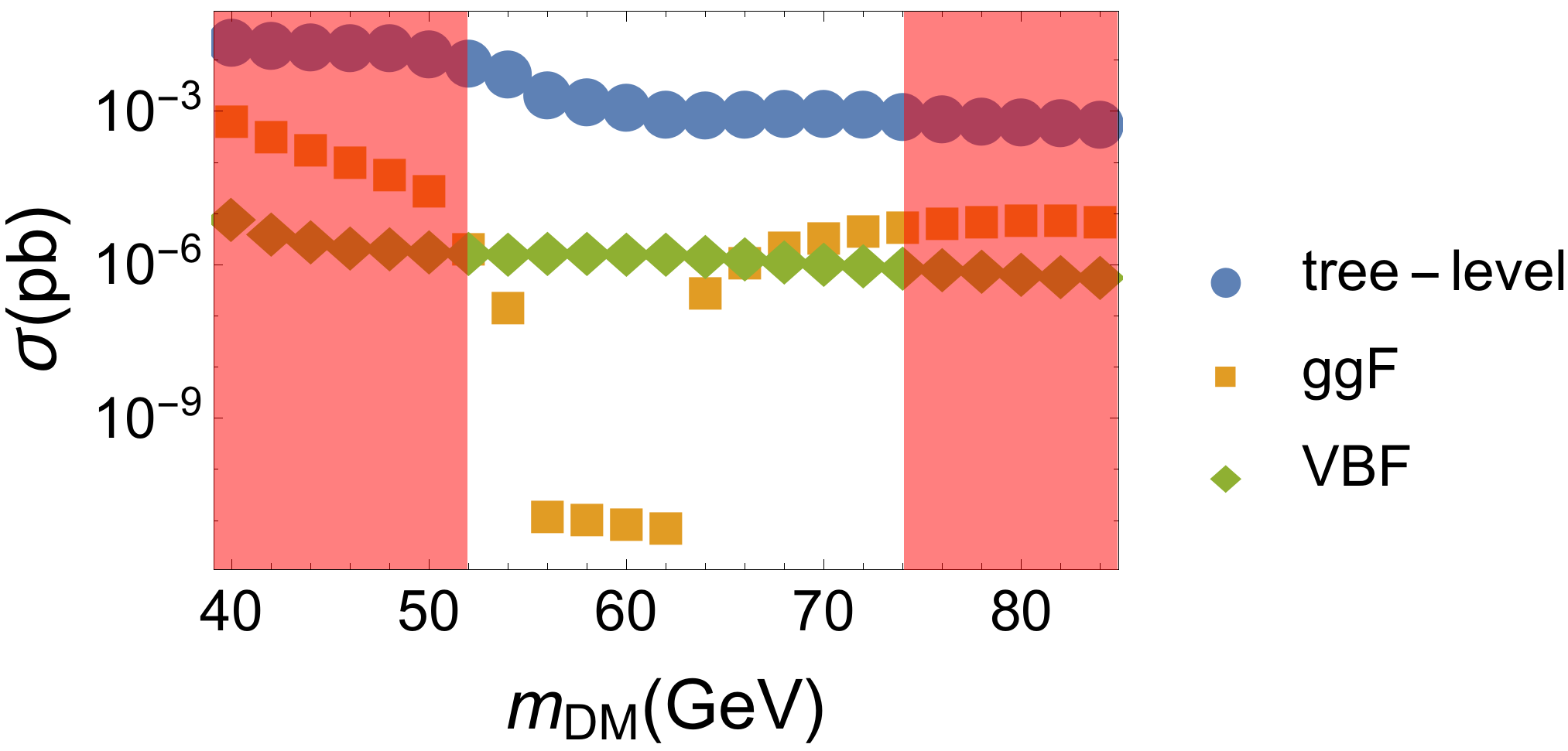}
\includegraphics[scale=0.3]{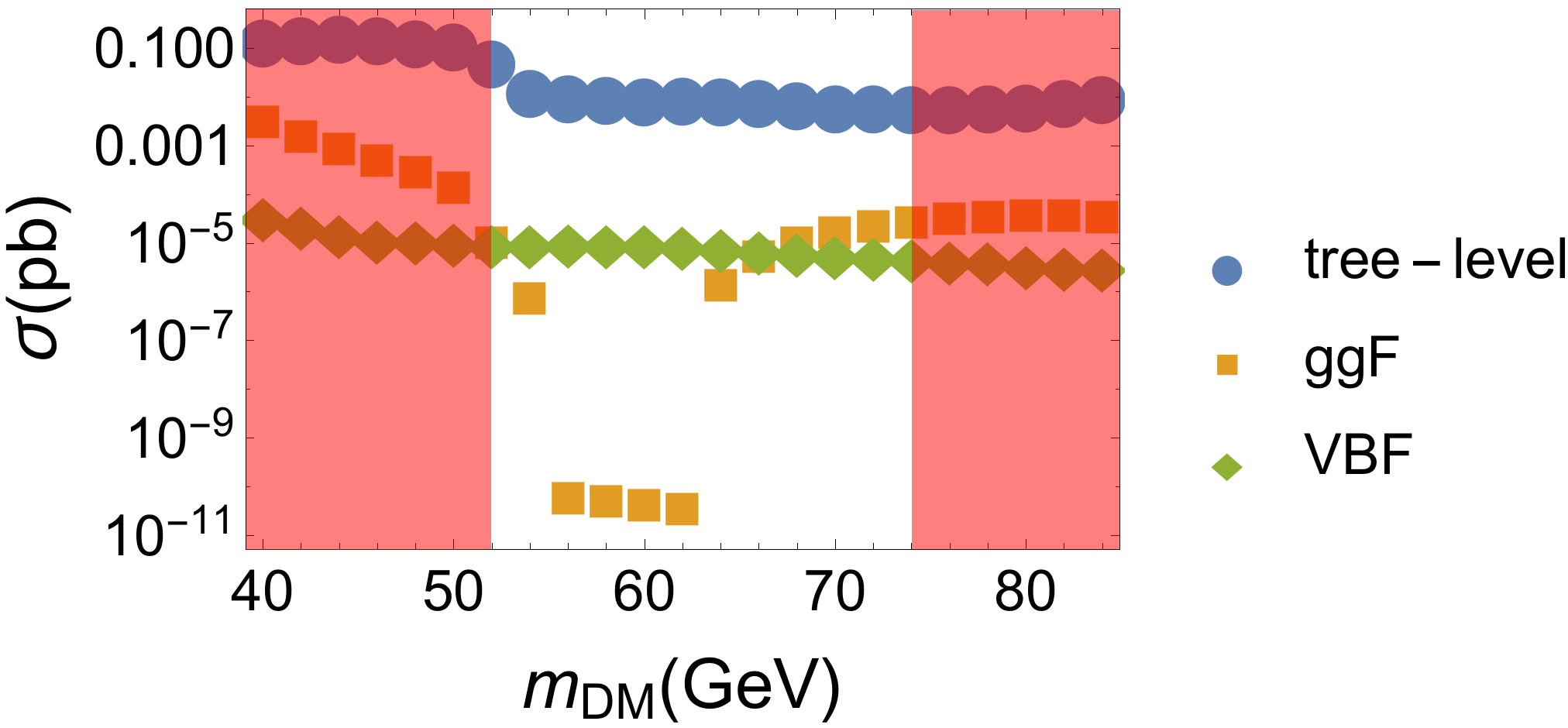}
\includegraphics[scale=0.8]{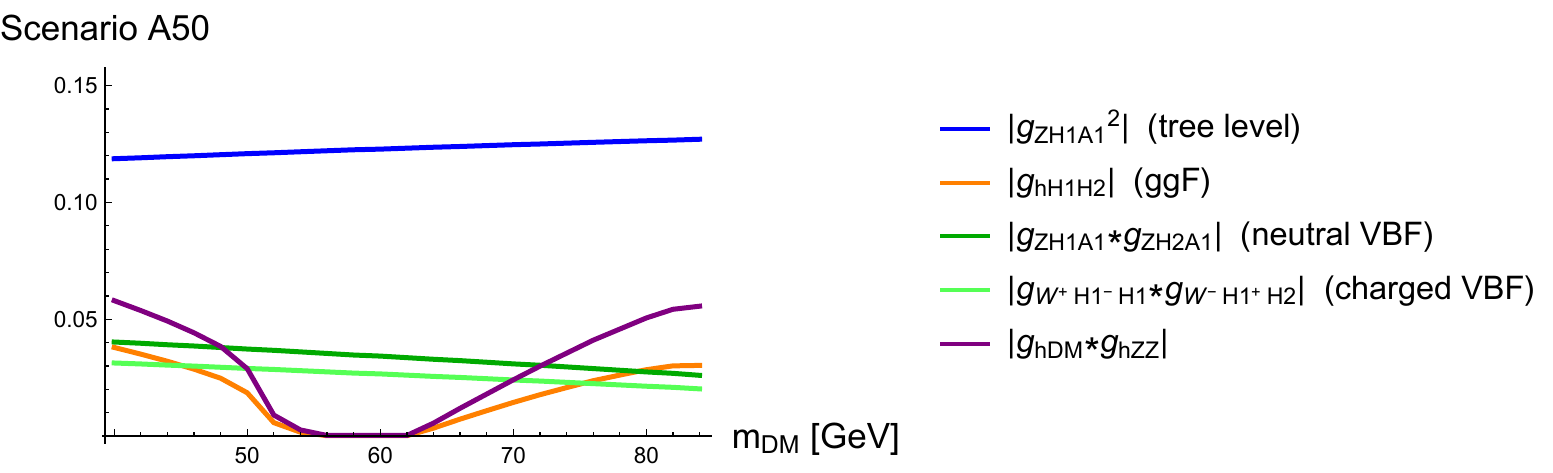}
\caption{The anatomy of scenario A50. The plots on the top show the cross-sections of the tree-level, ggF and VBF processes with leptonic (left) and hadronic (right) final states. The red regions are ruled out by LHC ($m_{DM} < 53$ GeV) and by direct detection ($m_{DM} > 73$ GeV). At the bottom we show the dominant couplings in each process with the same color coding where the Higgs-DM coupling is shown for reference. Note that the $g_{hH_1H_2}$ appears with the $K$-factor in the cross-section calculations.}
\label{A50-plot}
\end{center}
\end{figure}

\begin{figure}[h!]
\begin{center}
\includegraphics[scale=0.55]{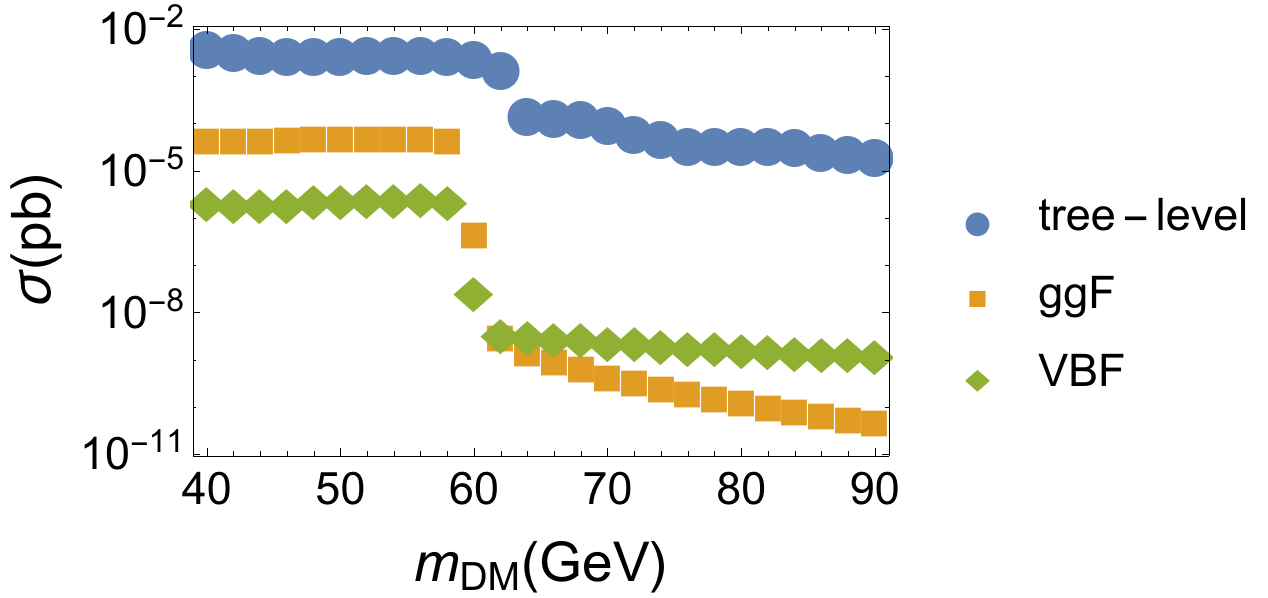}
\includegraphics[scale=0.55]{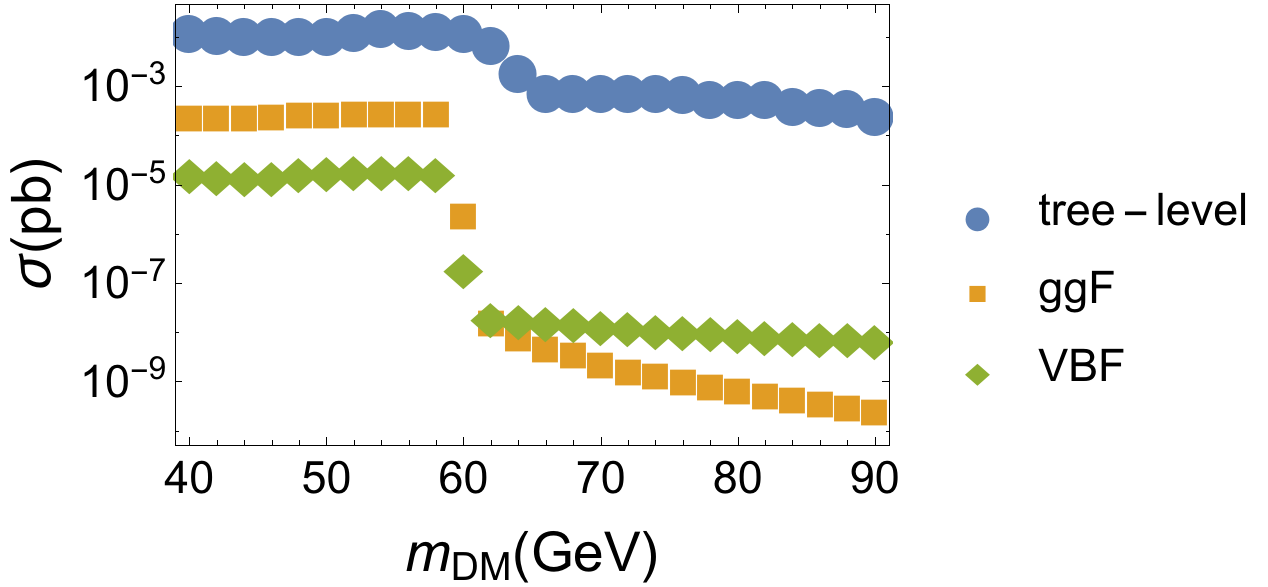}
\includegraphics[scale=0.8]{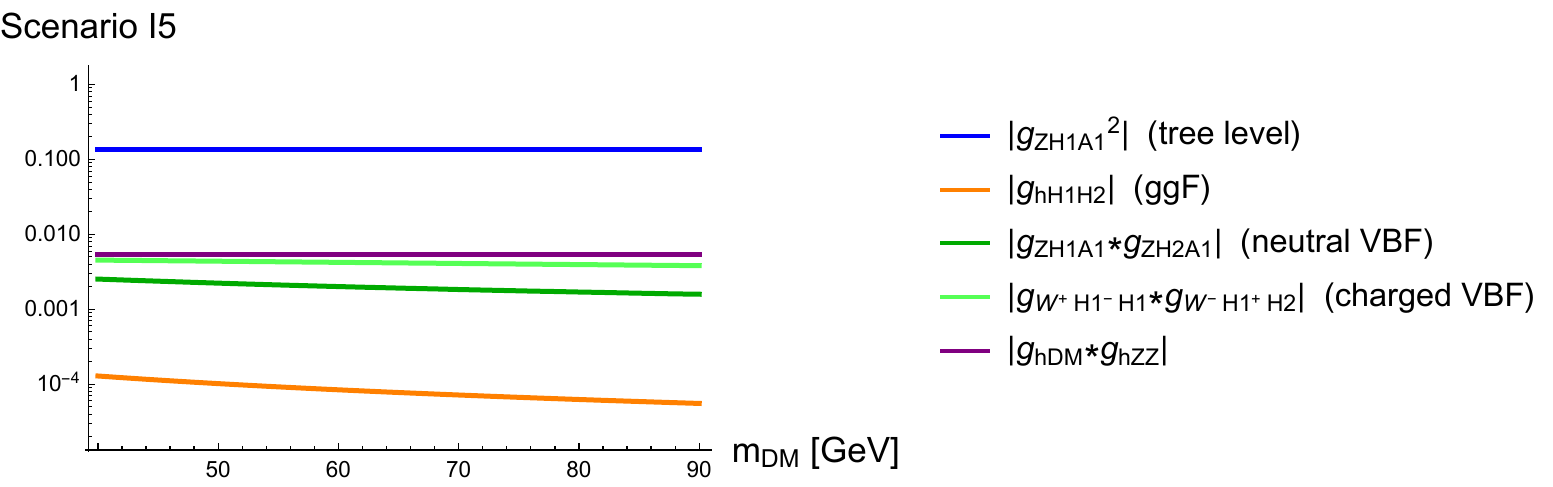}
\caption{The anatomy of scenario I5. The plots on the top show the cross-sections of the tree-level, ggF and VBF processes with leptonic (left) and hadronic (right) final states. At the bottom we show the dominant couplings in each process in Log scale with the same color coding where the Higgs-DM coupling is shown for reference. Note that the $g_{hH_1H_2}$ appears with the $K$-factor in the cross-section calculations.}
\label{I5-plot}
\end{center}
\end{figure}

\begin{figure}[h!]
\centering
\includegraphics[scale=0.55]{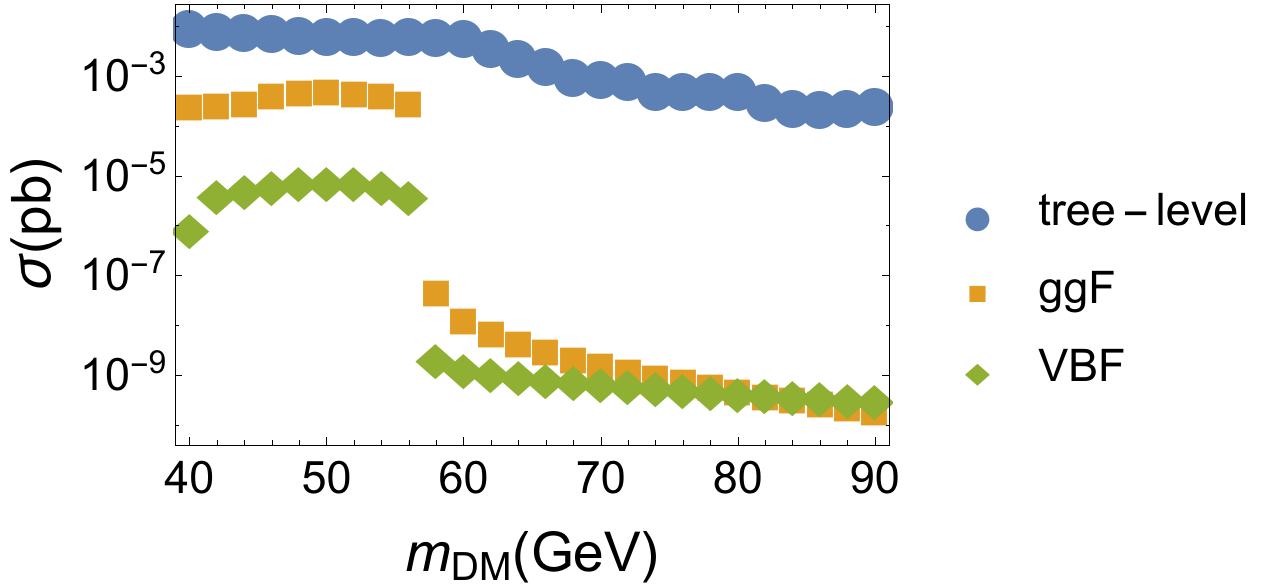}
\includegraphics[scale=0.55]{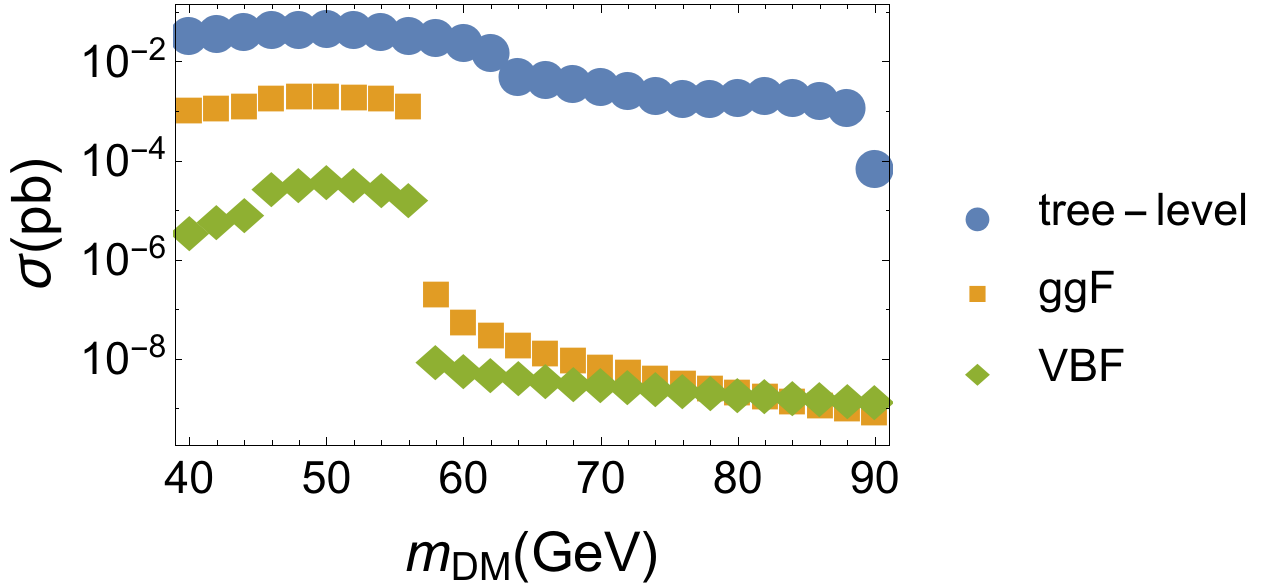}
\includegraphics[scale=0.8]{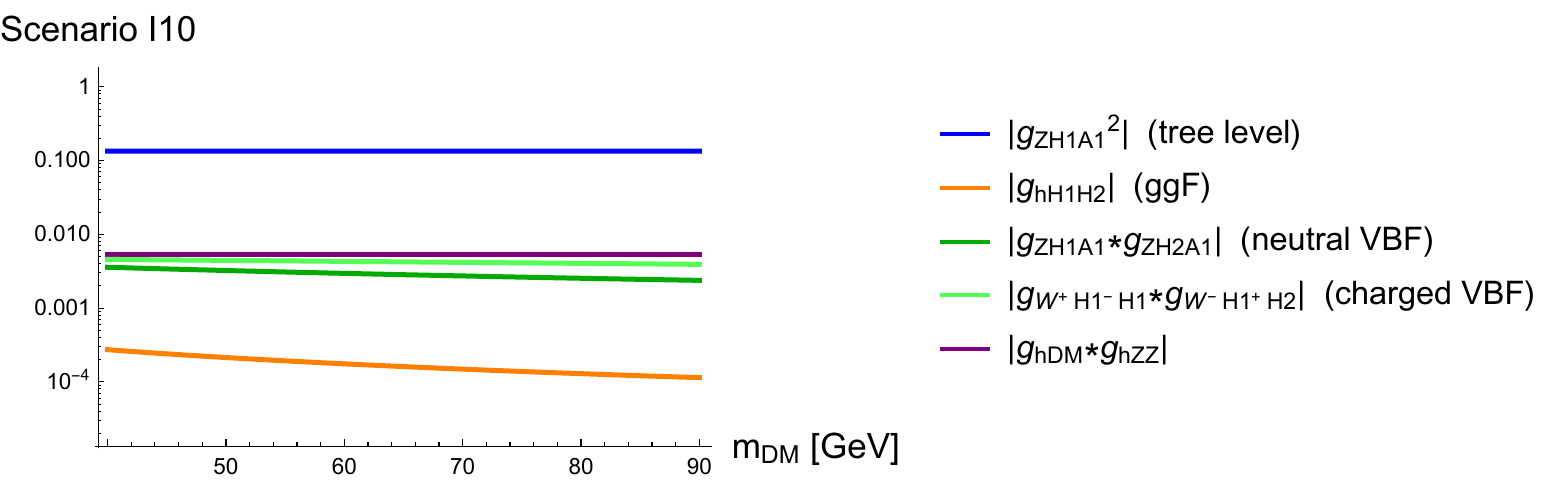}
\caption{The anatomy of scenario I10. The plots on the top show the cross-sections of the tree-level, ggF and VBF processes with leptonic (left) and hadronic (right) final states. At the bottom we show the dominant couplings in each process in Log scale with the same color coding where the Higgs-DM coupling is shown for reference. Note that the $g_{hH_1H_2}$ appears with the $K$-factor in the cross-section calculations.}
\label{I10-plot}
\end{figure}

Figs. \ref{A50-plot}--\ref{I10-plot}  show the anatomy of the given scenarios, which include not only
the cross-sections for leptonic ($\Et l^+ l^-$) and hadronic ($\Et q \bar q$) final states, but also the
relevant couplings in each case with the same colour coding. The Higgs-DM coupling is also shown for reference. 

For each benchmark scenario, we calculate the cross-section for three processes, namely, the ggF process (\ref{first}), the tree-level process (\ref{tree}) and the VBF process (\ref{VBF}) and present the dominant couplings entering in each case.

Let us first focus on scenario A50 presented in Fig. \ref{A50-plot}, which has
two special features. First, mass splittings between $H_1$ and other inert
particles are relatively large, as well as the main couplings  (in particular the $g_{ZH_1
A_1}$), which leads to large tree-level $Z$-mediated
cross-sections (the blue curve). Second, the Higgs-DM coupling, $g_{h H_1H_1}$,
is chosen such that the relic density is in exact agreement with Planck
measurements. To fulfil that, around the Higgs resonance the coupling
needs to be very small, of the order of $10^{-4}$ \cite{Keus:2014jha}. As the $g_{h
H_1 H_2}$ coupling is closely related to $g_{h H_1H_1}$, we observe  a
sudden dip for the orange curve ($g_{h H_1 H_2}$), which then leads to a
reduced cross-section for the ggF processes, driven by that particular
coupling. We also observe that the cross-section for the VBF processes, which
depends mainly on large mass splittings and relatively constant gauge
couplings, are as expected relatively constant for this benchmark.

Scenario I5, shown in Fig. \ref{I5-plot}, differs from the scenario A50 above. Here, the mass
splittings are much smaller, but also the Higgs-DM coupling is set to a
constant value for all masses, as seen in Fig.  \ref{I5-plot}. This makes the phase
space structure more visible. For $m_{H_1} < m_h/2$ all cross-sections are
roughly constant, with the ggF processes enhanced through the resonant
Higgs production. However, after crossing the Higgs resonance region, with
no increase of the Higgs-DM coupling to compensate for that, we observe a
rapid decrease of the value of the cross-section. For larger masses the cross-sections are too small to be observed for the current LHC luminosity.

Very similar behaviour is present for scenario I10 depicted in Fig. \ref{I10-plot}, where, similarly to scenario I5, the Higgs-DM coupling is set to a constant value for all masses.
Again we observe almost constant cross-sections, which are rapidly
reduced after we cross the Higgs threshold.



\section{Conclusions}
\label{summa}

In this paper, we have assessed the sensitivity of the LHC to Higgs signals in the   $\Et\; f \bar f$ channel, $f=u,d,c,s,b,e,\mu,\tau$, with invariant mass of the $ f \bar f$ pair much smaller than the
$Z$ mass. This signature would in fact point towards an underlying 3HDM structure of the Higgs sector, with one active and two inert doublets (so that the scenario can evocatively 
be nicknamed as I(2+1)HDM),  
induced by the decay $H_2\to H_1 f \bar f$, where $H_1$ represents the lightest CP-even neutral Higgs state from the inert sector (thereby being a DM candidate) and $H_2$ the next-to-lightest one. The decay proceeds via loop diagrams induced by the propagation of both SM weak gauge bosons ($W^\pm$ and $Z$) and inert Higgs states ($H^\pm_{1,2}$ and $A_{1,2}$) in  
two-, three- and four-point topologies, wherein the leading contribution comes from the intermediate decay step $H_2\to H_1\gamma^*$, involving a very low mass virtual photon scalarly polarised, eventually splitting in a collimated $f \bar f$ pair, which would be a distinctive signature of this Higgs construct. In fact, the corresponding 2HDM version, with one inert doublet only, i.e., the I(1+1)HDM, contains only one CP-even and only one CP-odd neutral Higgs state, so that no such a decay is possible owing to CP conservation. 

This signature would emerge from SM-like Higgs boson production, most copiously via ggF and VBF,  followed by a primary $h\to H_2H_1$ decay, so that the complete particle final state is $H_1 H_1f \bar f$, wherein the two DM candidates would produce missing transverse energy, accompanied by some hadronic activity in the forward and backward directions, originating by initial state gluon radiation or (anti)quark remnant jets, respectively, for ggF and VBF.  In fact, amongst the possible fermionic flavours $f$, the cleanest signature is afforded by the leptonic ones ($f=l$), in view of the overwhelming QCD background. While the muon and tauon cases are the cleanest, the latter being larger than the former (assuming only leptonic decays of the $\tau$'s), the electron case is potentially the one giving raise to 
the most spectacular signal, which,  
owing to parton distribution imbalances, so that the $h$ state would be boosted, would appear at 
detector level as a single EM shower with substantial $\Et$ surrounding it.

However, there is a substantial tree-level contribution, due to $q\bar q\to Z^{*}H_1H_1$ topologies (a first one involving single $h$-strahlung followed by $h\to H_1H_1$ splitting, a second one via a $Z^*Z^*H_1H_1$ vertex and a third one through $A_{1,2}H_1$ production followed by $A_{1,2}\to H_1 Z^*$ decay), which is potentially much larger than the aforementioned loop diagrams, thereby acting as an intrinsic background.  In fact, even though the $Z^*$ ought to be significantly off-shell in its transition to $f \bar f$ pairs to mimic the $\gamma^*\to f \bar f$ splitting, this can happen with substantial rates, because of the rather large value of the total $Z$ decay width. It is therefore clear that the $H_2\to H_1 f \bar f$ signal can only be established in presence of a rather small mass gap between $H_2$ and $H_1$. To this effect, we have then defined a few benchmarks on the I(2+1)HDM parameter space where the mass difference $m_{H_2}-m_{H_1}$ is taken to be increasingly small, varying from  50, to 10 to 5 GeV.
Correspondingly, we have seen  the relevance of the loop processes growing with respect to the tree-level one, with ggF dominating VBF,
to the point that {the former become comparable to the latter for cross-sections and BRs directly testable at Run 2 and/or Run 3 of the LHC. This is particularly true over 
  the DM mass region  observable at the CERN machine, i.e., 
for small values of the DM candidate mass, typically less than $m_h/2$. In this case, the cumulative signal can be almost within an order of magnitude or so of such an  intrinsic background.


We have obtained these results in the presence of up-to-date theoretical and experimental constraints, including amongst the latter those from colliders, DM searches and cosmological
relic density. Therefore, we believe that the advocated discovery channel might serve as smoking-gun (collider) signature of the I(2+1)HDM, that may enable one to distinguish it from the
I(1+1)HDM case, in a few years to come. In fact, once this signal is established and 
some knowledge of the $H_2$ and $H_1$ masses gained, the latter can be used to extract additional manifestations of the prevalent $H_2\to H_1\gamma^*$ decay, by considering the selection of additional splittings $\gamma^*\to f\bar f$, where $f$ can be identified with $q=u,d,c,s,b$, depending upon the relative value of $m_{H_2}-m_{H_1}$ and $2m_f$. Finally, in reaching these conclusions, we emphasise that we have done a complete one-loop calculation of the $H_2\to H_1 f\bar f$  decay process, including all topologies  entering through the same perturbative order, i.e., 
not only  those proceeding via $H_2\to H_1\gamma^*\to H_1 f\bar f$,
which was never attempted before, so that we have collected the relevant formulae in Ref. \cite{Cordero:2017owj}. 

In conclusion, the 3HDM with two inert doublets, provides a well motivated dark matter model with distinctive LHC signatures in certain
regions of parameter space arising from novel Higgs decays, the most spectacular being $e^+e^-+ \Et$ mono-shower.

\section*{Acknowledgements}
DR-C is supported by the Royal Society Newton International Fellowship NIF/R1/180813. SM, VK and DR-C are also partially supported by the H2020-MSCA-RISE-2014 grant no. 645722 (NonMinimalHiggs).
SM is financed in part through the NExT Institute. 
VK's research is partially supported by the Academy of Finland project 274503.
DS is supported in part by the National Science Center, Poland, through the HARMONIA
project under contract UMO-2015/18/M/ST2/00518.
JH-S and AC are supported by CONACYT (M\'exico), VIEP-BUAP and 
PRODEP-SEP (M\'exico) under the grant: ``Red Tem\'atica: F\'{\i}sica del Higgs y del Sabor".

\end{document}